\documentclass[aoas,preprint]{imsart}
\RequirePackage{natbib}

\usepackage{etex}
\reserveinserts{18}

\usepackage{amsmath,amsthm,amssymb,amsfonts}
\usepackage{amsxtra,amstext,latexsym,dsfont} % for \mathds{N}
\usepackage{mathtools}
\usepackage{wasysym}
\usepackage{stmaryrd}
\usepackage{yfonts}

\usepackage{parallel,accents}

\normalfont
\usepackage{bm}

\usepackage[pdftex]{graphicx}
\usepackage[dvipsnames]{xcolor}

\usepackage{pgf,tikz}
\usetikzlibrary{snakes}
\usetikzlibrary{backgrounds}
\usetikzlibrary{arrows}
\usetikzlibrary{patterns}
\usetikzlibrary{fit}
\usetikzlibrary{calc}
\usetikzlibrary{matrix}

\usepackage{lscape} 
\usepackage{array,longtable,booktabs,float,rotating}
\usepackage{dcolumn,ctable}

\usepackage[labelfont={bf},ruled,labelsep=period,small]{caption}

\usepackage[flushleft]{threeparttable}
\usepackage[linkcolor=black,colorlinks=true,citecolor=black,%ForestGreen,%pagebackref,%
bookmarks=true,bookmarksopen=true,bookmarksnumbered=true,bookmarksopenlevel=0,%
hyperindex=true,pdfpagemode=UseOutlines,linktocpage=true,pdfpagelayout=TwoColumnLeft,
pdftitle=Biostatistics,pdfstartview=FitH,pdffitwindow=true,urlcolor=blue]{hyperref}
\usepackage{ragged2e}
\usepackage{multirow,multicol}
\usepackage{fancyhdr,extramarks}
 \newcommand\sub{\@startsection%
     {subsubsection}{5}{0mm}{-1\baselineskip}{.01\baselineskip}%
     {\normalfont\itshape}}
 \renewcommand\subsubsection{\@startsection%
     {subsubsection}{3}{0mm}{-1\baselineskip}{.01\baselineskip}%
     {\normalfont\itshape}}
 \makeatother%--------------------------
\makeatletter
        \newcommand\Appendix[2][?]{%
            \refstepcounter{section}%
            \addcontentsline{toc}{appendix}%
                {\protect\numberline{\appendixname~\thesection}#1}%
            {\raggedleft\bfseries \appendixname\
                \thesection\par \centering#2\par}%
                \sectionmark{#1}%
                \@afterheading
                \addvspace{\baselineskip}}
        \newcommand\sAppendix[1]{%
            \raggedleft\bfseries\appendixname\par
            \@afterheading\addvspace{\baselineskip}}
    \makeatother

\normalsize \makeindex

\newcolumntype{A}{>{\centering}p{100pt}}
\newlength\savedwidth

\def\coldot{.}%
{\catcode`\.=\active%
    \gdef.{$\egroup\setbox2=\hbox to \dimen0 \bgroup$\coldot}}
\def\rightdots#1{%
    \setbox0=\hbox{$1$}\dimen0=#1\wd0%
    \setbox0=\hbox{$\coldot$}\advance\dimen0 \wd0%
    \setbox2=\hbox to \dimen0 {}%
    \setbox0=\hbox\bgroup\mathcode`\.="8000 $}
\def\endrightdots{$\hfil\egroup\box0\box2}

\newcolumntype{d}[1]{D{.}{.}{#1}}
\newcolumntype{A}{>{\centering}p{100pt}}
\newcolumntype{.}{D{.}{.}{-1}}
\newcolumntype{P}[2]{>{#1\raggedright\arraybackslash}p{#2}}

\DeclareFontFamily{U}{euc}{}% I chose euc because the chart is called Euler cursive
\DeclareFontShape{U}{euc}{m}{n}{<-6>eurm5<6-8>eurm7<8->eurm10}{}%
\newtheorem{thm}{Theorem}
\theoremstyle{plain}      
\theoremstyle{plain}      
\theoremstyle{plain}      
\theoremstyle{plain}      
\theoremstyle{definition} 
\theoremstyle{definition} 
\theoremstyle{definition} 
\theoremstyle{plain} \newtheorem{cor}{Corollary}
\theoremstyle{definition} 
\theoremstyle{plain} 
\theoremstyle{definition} 
\theoremstyle{definition} 
\theoremstyle{definition} 

\newcounter{nctr}
\newenvironment{en}{\begin{enumerate}}{\end{enumerate}}

\usepackage[rightbars,color]{changebar}
\cbcolor{blue}

\usepackage{url}
\usepackage{fancyvrb}

\newcommand\bb{\mathbb}

\newcommand\te{\text}

\newcommand\ca{\mathcal}

\newcommand\op{\operatorname}

\newcommand\argmin{\operatornamewithlimits{argmin}}

\newcommand\cov{\operatorname{\bb{C}ov}}

\newcommand\df{\te{df}}
\newcommand\E{\bb{E}}

\newcommand\iid{\op{iid}}

\newcommand\ind{\op{ind}}

\newcommand\lra{\longrightarrow}

\newcommand\lt{\left}

\newcommand\p{\bb{P}}           % Probability P
\newcommand\pri{^\prime}

\newcommand\rt{\right}

\newcommand\stack{\stackrel} % Over equal signs.

\newcommand\tth{^\text{th}}
\newcommand\var{\operatorname{\bb{V}ar}}

\newcommand\vech{\op{vech}}
\newcommand\wh{\widehat}

\newcommand\M{\bb{M}}  % Matrices
\newcommand\N{\bb{N}}  % Natural (counting numbers)
\newcommand\R{\bb{R}}  % Real
\newcommand\cL{\ca{L}} %
\newcommand\cX{\ca{X}} %
\newcommand\al{\alpha}

\newcommand\ep{\epsilon}

\newcommand\ka{\kappa}
\newcommand\la{\lambda}
\newcommand\sig{\sigma}

\newcommand\La{\Lambda}
\newcommand\Ga{\Gamma}

\newcommand\Sig{\Sigma}

\newcommand*\oline[1]{%
  \vbox{%
    \hrule height 0.5pt%                  % Line above with certain width
    \kern0.5ex%                          % Distance between line and content
    \hbox{%
      \kern-0.1em%                        % Distance between content and left side of box, negative values for lines shorter than content
      \ifmmode#1\else\ensuremath{#1}\fi%  % The content, typeset in dependence of mode
      \kern-0.1em%                        % Distance between content and left side of box, negative values for lines shorter than content
    }% end of hbox
  }% end of vbox
}

\startlocaldefs
\endlocaldefs

\begin{document}
\sloppy
%%%%%%%%%%%%%%%%%%%%%%%%%%%%%%%%%%%%%%%%%%%%%%%%%%%%%%%
\begin{frontmatter}
%%% Title of the paper
\title{Hypothesis Testing For Network Data \\ in Functional Neuroimaging}
\runtitle{Loss Functions for Matrix Means}

\begin{aug}
%%% Authors
\author{\fnms{Cedric} \snm{E.~Ginestet}\thanksref{t1,m1}\ead[label=e1]{cedric.ginestet@kcl.ac.uk}},
\author{\fnms{Jun} \snm{Li}\thanksref{t2,m2}\ead[label=e2]{junli14@bu.edu}},
\author{\fnms{Prakash} \snm{Balachandran}\thanksref{m2}\ead[label=e3]{prakashb@bu.edu}},
\author{\fnms{Steven} \snm{Rosenberg}\thanksref{t2,m2}\ead[label=e4]{sr@bu.edu}},
\and
\author{\fnms{Eric} \snm{D.~Kolaczyk}\thanksref{t1,t2,m2}\ead[label=e5]{kolaczyk@bu.edu}}

%%% Acknowledgements.
\thankstext{t1}{This work was supported by a grant from the Air Force
Office for Scientific Research (AFOSR), whose grant number is
FA9550-12-1-0102.} 
\thankstext{t2}{This material is also based upon work supported by,
or in part by, the U. S.~Army Research Laboratory and the U. S. Army
Research Office under contract/grant number W911NF1510440.}
\runauthor{C.E. Ginestet et al.}

%%% Universities
\affiliation{King's College London\thanksmark{m1} and Boston University\thanksmark{m2}}

%%% Full addresses:
\address{Department of Biostatistics\\ and Health Informatics,\\
Institute of Psychiatry, \\Psychology and Neuroscience, \\
King's College London \\
London SE5 8AF. \\
\printead{e1}}

\address{Department of Mathematics and Statistics, \\
Boston University,\\
Boston, MA 02215. \\
\printead{e2}\\
\phantom{E-mail:\ }\printead*{e3}\\
\phantom{E-mail:\ }\printead*{e4}\\
\phantom{E-mail:\ }\printead*{e5}}
\end{aug}

%%%%%%%%%%%%%%%%%%%%%%%%%%%%%%%%%%%%%%%%%%%%%%%%%%%%%%%
\begin{abstract}
In recent years, it has become common practice in neuroscience to use networks 
to summarize relational information in a set of measurements,
typically assumed to be reflective of either functional or structural
relationships between regions of interest in the brain.  
One of the most basic tasks of interest in the analysis of such data
is the testing of hypotheses, in answer to questions
such as ```Is there a difference between the networks of
these two groups of subjects?''  In the classical setting,
where the unit of interest is a scalar or a vector, such questions are
answered through the use of familiar two-sample testing
strategies. Networks, however, are not Euclidean objects, and hence
classical methods do not directly apply. We address this
challenge by drawing on concepts and techniques from geometry, and
high-dimensional statistical inference. Our work is based on a precise geometric
characterization of the space of graph Laplacian matrices and a
nonparametric notion of averaging due to Fr{\'e}chet.  We motivate and
illustrate our resulting methodologies for testing in the context of
networks derived from functional neuroimaging data on human subjects
from the 1000 Functional Connectomes Project. In particular, we show
that this global test is more statistical powerful, than a
mass-univariate approach. In addition, we have also provided a method
for visualizing the individual contribution of each edge to the
overall test statistic.
\end{abstract}
\end{frontmatter}

%%%%%%%%%%%%%%%%%%%%%%%%%%%%%%%%%%%%%%%%%%%%%%%%%
%%%%%%%%%%%%%%%%%%%%%%%%%%%%%%%%%%%%%%%%%%%%%%%%%
%%%%%%%%%%%%%%%%%%%%%%%%%%%%%%%%%%%%%%%%%%%%%%%%%
\section{Introduction}\label{sec:introduction}
Functional neuroimaging data has been central to the advancement of
our understanding of the human brain. Neuroimaging data sets are
increasingly approached from a graph-theoretical perspective, using
the tools of modern network science \citep{Bullmore2009}. This has
elicited the interest of statisticians working in that area.   At the level of
basic measurements, neuroimaging data can be said to consist typically 
of a set of signals (usually time series) at each of a collection of pixels
(in two dimensions) or voxels (in three dimensions).  Building from such data,
various forms of higher-level data representations are employed in
neuroimaging.  Traditionally,
two- and three-dimensional images have, naturally, been the norm, but increasingly
in recent years there has emerged a substantial interest in network-based representations.

%%%%%%%%%%%%%%%%%%%%%%%%%%%%%%%%%%%%%%%%%%%%%%%%%
\subsection{Motivation}\label{sec:motivation}
Let $G=(V,E)$ denote a graph, based on $d=|V|$ vertices.  In this
setting, the vertices $v\in V$
correspond to regions of interest (ROIs) in the brain, often
pre-defined through considerations
of the underlying neurobiology (e.g., the putamen or the cuneus).
Edges $\{u,v\}\in E$ between 
vertices $u$ and $v$ are used to denote a measure of association
between the corresponding ROIs.
Depending on the imaging modality used, the notion of `association'
may vary.  For example, in diffusion tensor imaging (DTI),
associations are taken to be representative of structural connectivity
between brain regions.  On the other hand, in functional magnetic
resonance imaging (fMRI), associations
are instead thought to represent functional connectivity, in the sense
that the two regions of the brain
participate  together in the achievement of some higher-order
function, often in the context of performing
some task (e.g., counting from $1$ to $10$).

With neuroimaging now a standard tool in clinical neuroscience, and
with the advent of several
major neuroscience research initiatives -- perhaps most prominent
being the recently announced
Brain Research Accelerated by Innovative Neurotechnologies (BRAIN)
initiative -- we are quickly
moving towards a time in which we will have available databases composed of 
large collections of secondary data in the form of network-based data objects.  
Faced with databases in which networks are a fundamental unit of data, it will be necessary to have in place
the statistical tools to answer such questions as, ``What  is the `average' of a collection of networks?'' and
``Do these networks differ, on average, from a given
nominal network?,'' as well as ``Do two collections of networks differ on average?'' and  ``What factors (e.g., age, gender, etc.)
appear to contribute to differences in networks?'', or finally, say, ``Has there been a change in the networks for a given
subpopulation from yesterday to today?''  In order to answer these and similar questions, we require network-based analogues of
classical tools for statistical estimation and hypothesis testing.

While these classical tools are among the most fundamental and ubiquitous in use in practice,
their extension to network-based datasets, however, is not immediate and, in fact, can
be expected to be highly non-trivial.  The main challenge in such an
extension is due to the simple fact that networks are not Euclidean objects
(for which classical methods were developed) 
-- rather, they are combinatorial objects, defined simply through their sets of vertices and edges.
Nevertheless, our work here in this paper demonstrates that networks can be associated
with certain natural subsets of Euclidean space, and furthermore demonstrates that through a combination of
tools from geometry, probability on manifolds, and high-dimensional statistical analysis it is possible to develop 
a principled and practical framework in analogy to classical tools.  In particular, we focus on the development of
an asymptotic framework for one- and two-sample hypothesis testing.

Key to our approach is the correspondence between an undirected graph $G$ and its Laplacian,
where the latter is defined as the matrix $L=D-W$; with $W$ denoting the $d\times d$ adjacency matrix of $G$
and $D$ a diagonal matrix with the vertex degrees along the diagonal.  When $G$ has no self-loops and
no multi-edges, the correspondence between graphs $G$ and Laplacians $L$ is one-to-one.  Our 
work takes place in the space of graph Laplacians.  Importantly, this requires working not in 
standard Euclidean space $\mathbb R^n$, but rather on certain
subsets of Euclidean space which are either submanifolds of $\mathbb
R^n$, or submanifolds of $\mathbb R^n$ with corners.
While these subsets of Euclidean space have the potential to be complicated in nature, we show that in the absence of
any nontrivial structural constraints on the graphs $G$, the geometry of these subsets is sufficiently `nice' to allow
for a straightfoward definition of distance between networks to emerge.  

Our goal in this work is the development of one- and two-sample tests for network data objects 
that rely on a certain sense of `average'.  We adopt the concept of Fr{\'e}chet means in defining 
what average signifies in our context.  Recall that, for a metric
space, $(\cX,\rho)$, and a probability measure, $Q$,
on its Borel $\sigma$-field, under appropriate conditions,
the Fr{\'e}chet mean of $Q$ is defined as the (possibly nonunique) minimizer
\begin{equation}
\mu := \argmin_{x\in\cX} \int\limits_{\cX}\rho^{2}(x,y) Q(dy).
\label{eq:Frechet.mean}
\end{equation}
Similarly, for any sample of realizations from $Q$ on $\cX$, denoted
$Y:=\{Y_{1},\ldots,Y_{n}\}$, the corresponding sample Fr{\'e}chet mean is defined as
\begin{equation}
    \wh{\mu}_{n}(Y) := \argmin_{x\in\cX}\frac{1}{n}\sum_{i=1}^{n}\rho^{2}(x,Y_{i}). 
\label{eq:sample.Frechet.mean}
\end{equation}
Thus, the distance $\rho$ that emerges from our study of the geometry
of the space of networks implicitly defines a corresponding notion of
how to `average' networks.

Drawing on results from nonparametric statistical inference on manifolds, we are then 
able to establish a central limit theory for such averages and, in
turn, construct the asymptotic 
distributions of natural analogues of one- and two-sample $z$-tests.
These tests require knowledge of the covariance
among the edges of our networks, which can be expected to be
unavailable in practice. Nevertheless, we show how
recent advances in the estimation of large, structured covariance
matrices can be fruitfully brought to bear in our context, and provide 
researchers with greater statistical power than a mass-univariate
approach, which is the standard approach in this field. 

%%%%%%%%%%%%%%%%%%%%%%%%%%%%%%%%%%%%
\subsection{The 1000 Functional Connectomes Project}\label{sec:bg.neuro}
Our approach is motivated by and illustrated with data from the 1000
Functional Connectomes Project (FCP). This major MRI data-sharing
initiative was launched in 2010 \citep{Biswal2010}. 
The impetus for the 1000 FCP was given by a need to make widely
accessible neuroimaging data, which are costly and time-consuming to
collect \citep{Biswal2010}. This was conducted within the so-called
``discovery science'' paradigm, paralleling similar initiatives in
systems biology. The 1000 FCP constituted the largest data
set of its kind, at the time of its release. As for the use of such
large data sets in genetics, it is believed that facilitating access
to high-throughput data generates economies of scale that are likely
to lead to more numerous and more substantive research findings. 

%%% MRI, physics details, scanners, centers:
The 1000 FCP describes functional neuroimaging data from 1093
subjects, located in 24 community-based centers. The mean age of the
participants is 29 years, and all subjects were 18 years-old or
older. Each individual scan lasted between 2.2 and 20 minutes. The
strength of the MRI scanner varied across centers, with $n=970$
scans at 3T and $n=123$ at 1.5T. Voxel-size was 1.5--5mm within the
plane; and slice thickness was 3--8mm. The ethics committee in each
contributing data center approved the project; and the institutional
review boards of the NYU Langone Medical Center and of the New Jersey
Medical School approved the dissemination of the data. 
This freely available data set has been extensively used in the
neuroimaging literature \citep{Yan2013,Tomasi2010,Zuo2012}.

%%% Network constructions: Atlas, Correlation and mutual information (metrics).
The individual fMRI scans were parcellated into a set of 50 cortical and subcortical
regions, using the Automated Anatomical Labeling (AAL) template
\citep{Tzourio-Mazoyer2002}. Note, that that the resulting
connectivity networks are sensitive to our particular choice of
parcellation, and that the results in this paper need not
generalize to other templates \citep[see][for a review]{Wang2009}.
The voxel-specific time series in each of
these regions were aggregated to form mean regional time series, as
commonly done in the study of the human connectome
\citep[see for example][]{Achard2006}. The resulting regional time series
were then compared using two different measures of association. We
here considered the correlation coefficient since this measure has
proved to be popular in the neuroimaging literature
\citep{Ginestet2011a,Pachou2008,Micheloyannis2009}. 

\begin{figure}[t]
   \raggedright
   %%% Two simulated structures:
   \hspace{2cm}\textbf{(A) Sex} \hspace{5.3cm} \textbf{(B) Age}\\
   \vspace{.25cm}
   \hspace{1.25cm}\textit{Female} \hspace{1cm} \textit{Male}\hspace{2.65cm}
   $x\leq 22$ \hspace{.6cm} $22< x \leq 32$ \hspace{.52cm} $32<x$\\
   \vspace{0.10cm}
   %%% Sex Laplacians.
   \hspace{1.0cm}
   \includegraphics[width=1.5cm]{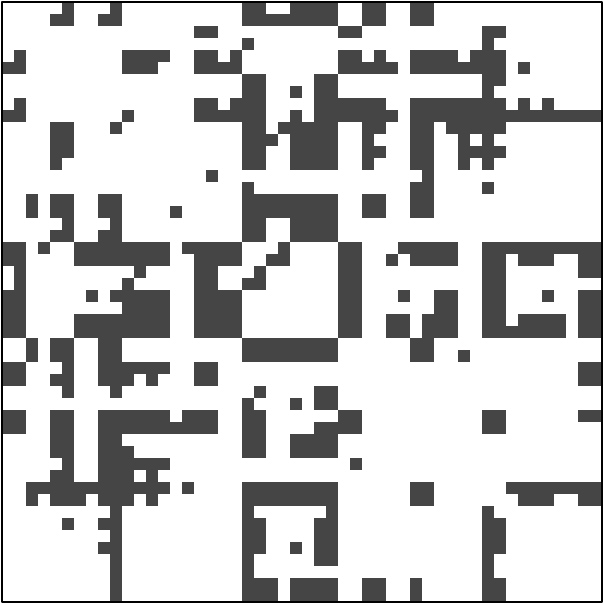}
   \hspace{.25cm}
   \includegraphics[width=1.5cm]{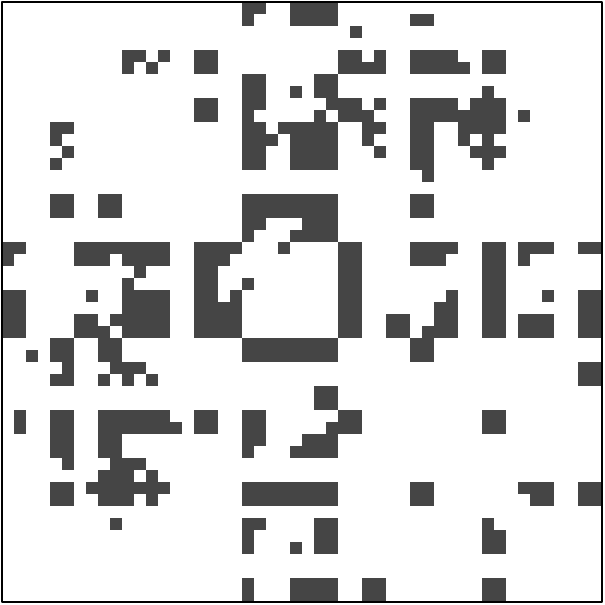}
   %%% Age Laplacians.
   \hspace{1.75cm}
   \includegraphics[width=1.5cm]{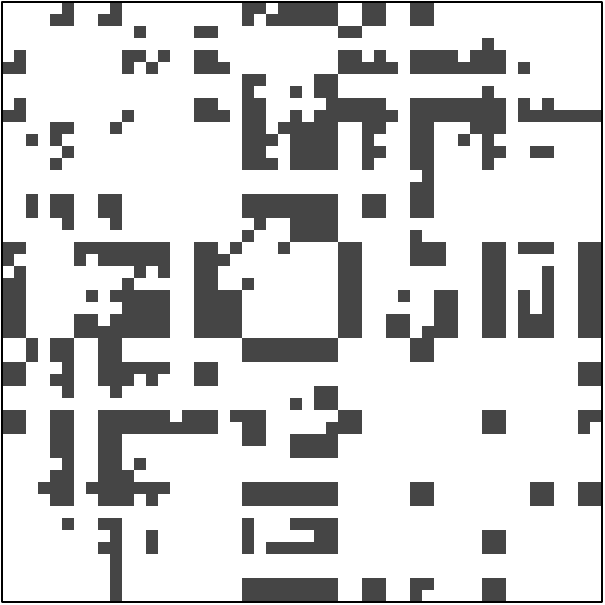}
   \hspace{.25cm}
   \includegraphics[width=1.5cm]{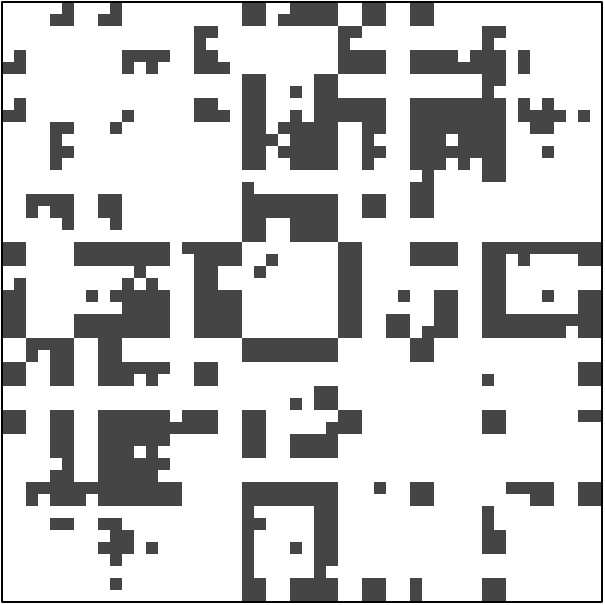}
   \hspace{.25cm}
   \includegraphics[width=1.5cm]{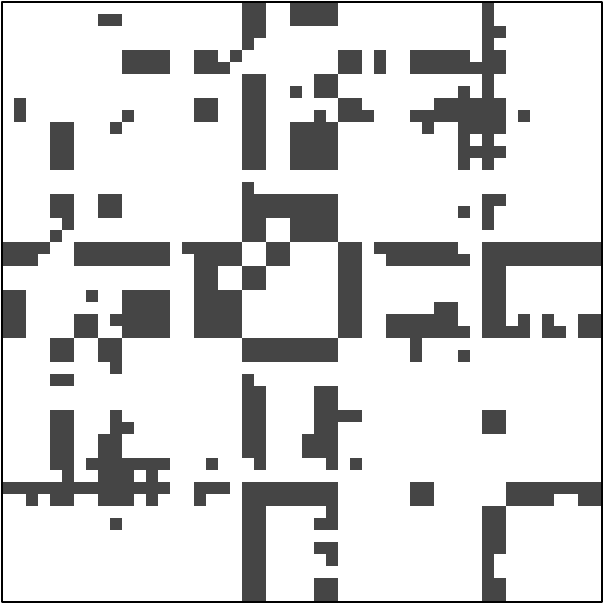}
   \caption{Descriptive statistics for the 1000 FCP data set. In panel
     (A), the group-specific mean
     Laplacians for networks over 50 AAL vertices in females
     ($n_{f}=555$), and males ($n_{m}=462$). Similarly, in panel
     (B), the age variable has been divided into three groups
     ($n_{1}=386$, $n_{2}=297$, and $n_{3}=334$), and the
     respective means are reported for each age group. The
     Laplacians have been binarized with respect to the $75\tth$
     percentile of the overall distribution of the entries in the full
     1000 FCP database. (Black indicates entries greater or equal
     than that percentile). 
     \label{fig:sexage}}
\end{figure}

%%% Experimental Design: notion of averages and GLOBAL comparisons.
Subjects in the 1000 FCP data can be subdivided with respect to
sex. Several groups of researchers have previously considered the
impact of sex differences on resting-state connectivity 
\citep{Biswal2010,Tomasi2011}. It is hypothesized that
sexual dimorphism in human genomic expression is likely to affect a
wide range of physiological variables \citep{Ellegren2007}. In
particular, differences in hormonal profiles (e.g.~estrogen) during
brain development are known to be related to region-specific effects 
\citep{McEwen1999}. Thus, it is of interest to compare the subject-specific
networks of males and females in the 1000 FCP data set (see Figure
\ref{fig:sexage}). Observe that 
previous research in this field has established \textit{local} sex
differences in connectivity by considering individual edge weights
\citep{Biswal2010,Tomasi2011}. By contrast, we are here investigating
the effect of sex differences on \textit{entire} networks. 

It is here useful to distinguish between these two types of network
data analysis in neuroimaging. While local analysis focuses on edge-specific
statistics; global analysis instead considers network topological
properties such as the shortest-path length. In this paper, we are
extending the latter by providing a framework for identifying the mean
network, and characterizing the space of all possible such
networks. 

% \subsection{Structure of the Paper}\label{sec:structure}
The organization of this paper is as follows. In Section \ref{sec:bg},
we describe the statistical and mathematical background of this type
of research questions. In Section \ref{sec:char.nets}, we provide
a geometrical characterization of the space of networks under
scrutiny. In Section \ref{sec:inference}, we describe how certain
central limit theorems can be adapted to this space, in order to construct a
statistical inferential framework for network data. A simulation study
exploring the relationship between statistical power and various
aspects of neuroimaging data is reported in Section \ref{sec:sims}.
In Section \ref{sec:data}, we apply this framework to the analysis of
a subset of the data from the 1000 FCP. These results and the
potential extensions of the proposed statistical tests are then
discussed in Section \ref{sec:discussion}.

%%%%%%%%%%%%%%%%%%%%%%
%%%%%%%%%%%%%%%%%%%%%%

%%%%%%%%%%%%%%%%%%%%%%%%%%%%%%%%%%%%%%%%%%%%%%%%%%
\section{Related Work}\label{sec:bg}
% \subsection{The 1000 Functional Connectomes Project (FCP)}
% \label{sec:bg.neuro}

% The work at hand is located at the nexus of several disciplines.
At the heart of the class of statistical problems we wish to address
is a desire to summarize and compare groups of network data objects
in a statistically principled manner.  There are, of course, already a
variety of numerical devices available for carrying out certain descriptive summaries
and comparisons. Basic set-theoretic operations (e.g., union, intersection,
symmetric difference) are all well-defined for graphs. More
broadly, various metrics, such as the Hamming distance, have been borrowed
from other fields and applied to graphs. Currently, the mainstay in the analysis
of network data in neuroimaging, is the mass-univariate approach in
which independent tests are conducted for every edge, adjusting for
multiple testing. See
\citet{Ginestet2014} for a survey of such methods in the
context of functional neuroimaging. 

Such mass-univariate approaches, however, fail to draw
inference about networks as a whole. In particular, it is unclear
whether multiple local differences necessarily lead to globally
significant differences. One may tackle this problem by treating
network data objects as data points.
What is lacking to achieve this, however, is the necessary
mathematical foundation -- establishing a formal `space' of graphs,
equipped with a formal metric, with understood geometric and
topological properties, so that a formal notion of probability and
measure can be defined, all underlying the desired theory and methods
for the hypothesis testing problems of interest here.

Networks are not the only data type for which standard Euclidean-based
methods are insufficient.  Statistical inference on manifolds --
in particular on spheres and shapes spaces -- has a fairly long history.
There is a substantial literature on statistics on spheres,
or so-called directional statistics,
going back to a seminal paper by R.A. Fisher in 1953 \citep{Fisher1953},
and works by \citet{Watson1983}, \citet{Mardia2000}, and \citet{Fisher1987},
among others. Statistical analysis on shapes that are landmark-based
was pioneered by \citet{Kendall1977}, \citet{Kendall1984} and \citet{Bookstein1980}.
Inference in these settings takes various forms. Nonparametric forms of inference
typically employ a notion of averaging due to 
\citet{Frechet1948}, as we do in this paper.  
Nevertheless, little work has been pursued with manifolds given as
some general metric space -- such as the spaces of networks that are
our main interest. The most related work seems to be  due to
\citet{Billera2001} and \citet{Barden2013}, who study the metric
geometry of the space of phylogenetic trees and derive a central limit
theorem for the Fr\'echet mean in such spaces.  Also see the related work
of Marron and colleagues in the context of so-called object-oriented
data analysis with trees \citep{Wang2007,Aydin2009}. 

In order to establish a formal characterization of a well-defined `space' of
networks, it is natural to associate a network with a matrix. And, while
there are several such matrices that might be used, we have found that
the (combinatoral) graph Laplacian is particularly appropriate.  
The Laplacian falls in the cone of symmetric positive (semi)definite
(PSD) matrices.  A substantial amount of effort has been expended on uncovering the
mathematical properties of the PSD cone
\citep{Bhatia1997,Moakher2011}.
In addition, there has in recent years
been quite a lot of work exploring the various notions of `average'
induced upon this manifold by the underlying choices of geometry
\citep{Arsigny2007,Moakher2005,Bonnabel2009}.
Finally, depending on the choice of average adopted,
there are results establishing the probabilistic and statistical properties
of averages through CLTs \citep{Bhattacharya2003,Bhattacharya2005,
Bhattacharya2012,Kendall2011}. Much of this research has been
motivated by shape analysis \citep{Le2000,Le2001}, but many of these
results have been developed in other areas of applications where matrices
play a key role such as in DTI \citep{Dryden2009}.

%\lizhen{Note: need to add the references on SPDS in the literature of
%diffusion tensor imaging}

However, the space of graph Laplacians forms
a \emph{subset} of the PSD cone and, furthermore, by definition this subset
intersects in a non-trivial fashion with the boundary of this
cone.  Therefore, results for PSD matrices do not carry over immediately
to the space of graph Laplacians -- the latter must necessarily be
studied in its own right.  At present, while graph Laplacians
as individual objects are well-studied --see \citet{Chung1997},
who discusses discrete eigenvalue and isoperimetric estimates
analogous to Riemannian estimates \citep[see also][]{Chavel, Xia} --
there appears to be no formal body of results to date establishing
the properties of the \emph{space} of graph Laplacians -- and certainly
none that reflects the impact of what have become established
canonical properties of complex networks (e.g., sparseness, small-world, etc.).
The closest work of which we are aware is, for example, recent work in
the signal processing literature, characterizing subspaces of the PSD
cone corresponding to subsets of covariance matrices sharing certain
simple structural properties such as rank or trace constraints
\citep{Krishnamachari2013a}.

%%%%%% Input from SR
A certain notion of embedding is crucial to the mathematical and
probabilistic theory underlying our approach.  There are,
in fact, different uses of the term ``embedding''.  Our work involves
averaging or comparing different networks/graphs via the distance
between network Laplacians  computed 
by first embedding (i.e. smoothly injecting) the set of Laplacian
matrices into a Euclidean space; here ``embedding" is defined 
as in the differentiable topology literature \citep[see
chap.~7 in][]{Lee2006}. 
This seems to have advantages over comparing networks via
e.g. isometric embeddings of the graph itself into $\mathbb R^3$,
for which computation of the types of distance functions that have
been useful (e.g.~Gromov-Hausdorff distance) is impractical.

In addition, there is also the large literature on graph embedding,
which maps a graph onto a typically low-dimensional Euclidean space
using eigenvector/eigenvalue information of the adjacency matrix or
associated Laplacian
\citep{linial1995geometry,Linial2002,yan2007graph,fu2013graph}.
Graph embedding methods are  very different from differentiable topology
techniques. In particular,  the image of a  graph embedding is often
used as a dimension-reduction tool.  This map in general has some
distortion, and so is not an isometry. This change in the
geometry from the domain space to the range space implies that
the precise inference framework for manifolds that we employ here, as
described below, cannot be applied to graph embeddings.  Thus, 
there is no natural notion of average and projection onto the image
under a graph embedding, and in fact such a projection may not
exist. On the other hand, our notion of embedding, which considers
the spaces of Laplacians as a manifold,  does not reduce dimension,
preserves all the raw information in a specific graph, and allows
analysis of averages and projections by geometric methods.  
 
%%%%%%%%%%%%%%%%%%%%%%%%%%%%%%%%%%%%%%%%%%%%%%%%%
%%%%%%%%%%%%%%%%%%%%%%%%%%%%%%%%%%%%%%%%%%%%%%%%%
%%%%%%%%%%%%%%%%%%%%%%%%%%%%%%%%%%%%%%%%%%%%%%%%%
\section{Characterization of Spaces of Networks}\label{sec:char.nets}
In this section, we establish the necessary mathematical properties associated with
a certain notion of a `space' of networks, from which a natural notion of
`averaging' emerges.  In fact, we offer several variations of a space
of networks and, in doing so,
illustrate how even relatively simple constraints on network topology
affect the geometry of these spaces.
The geometry is important when seeking to develop the corresponding
probabilistic behavior 
of averages of networks, as we do in Section \ref{sec:inference},
which also informs the sampling distributions
of the one- and two-sample test statistics that we develop.

\subsection{Main Results}\label{sec:S general}
Let $G=(V,E,W)$ be a \emph{weighted} undirected graph,
for weights $w_{ij}=w_{ji}\ge 0$, where equality with zero holds if and only
if $\{i,j\}\notin E$.  Assume $G$ to be simple (i.e., no self-loops
or multi-edges).  We associate uniquely with each graph $G$ its
graph Laplacian $L=D(W)-W$, where $D$ is a diagonal matrix of
weighted degrees (also called vertex strengths), i.e.,
$D_{jj} = d_j(W) = \sum_{i\ne j} w_{ij}$.  We further assume in most of what follows 
that $G$ is connected, in which case $L$ has one (and only one) zero eigenvalue
and all the others are positive (and hence $L$ is positive semi-definite).  

Under the assumption that $G$ is simple, there is a one-to-one correspondence between
a graph $G$ and its Laplacian matrix $L$.  We therefore define our
space of networks using the corresponding space of Laplacians. In the
following theorem, we 
show that an initial notion of the space of graph Laplacians
over $d$ nodes admits a relatively simple topology, which can be
described as a convex subset of an affine space in $\R^{d^{2}}$. 
\begin{thm}\label{thm:S}
   The set $\cL_{d}$ of $d\times d$ matrices $A$, satisfying: 
   \begin{en}
     \item[(1)] $\op{Rank}(A)=d-1$,
     \item[(2)] Symmetry, $A\pri=A$,
     \item[(3)] Positive semi-definiteness, $A\geq0$,
     \item[(4)] The entries in each row sum to 0,
     \item[(5)] The off-diagonal entries are negative, $a_{ij}<0$;
   \end{en}
   forms a submanifold of $\R^{d^{2}}$ of dimension $d(d-1)/2$. In fact, 
   $\cL_{d}$ is a convex subset of an affine space in
   $\R^{d^{2}}$ of dimension $d(d-1)/2$. 
\end{thm} 

A proof of this theorem is in the Supplementary Material.
The practical importance of this result is that
$\cL_d$ admits several Riemannian metrics \citep{Arsigny2007}, which give rise to a
restricted class of distance functions. For example, any one of these
metrics turns $\cL_d$ into a length space in the sense of
\citet{Gromov2001}, i.e. the distance between any two points  $A, B\in
\cL_d$ is the length of some path from $A$ to $B$. Also, all
the usual notions of curvature, and its influence on variations of
geodesics, come into play.
%%%%%%%%%%%%%%%

However, we note that the definition of $\cL_{d}$ requires that
\emph{every} potential edge in $G$
be present, with edges only distinguishable in terms of the relative magnitude
of their weights.  Consider the 
description of the 1000 FCP data in Section~\ref{sec:bg.neuro}.  For
the case where our network is defined to be, say,
the matrix $W$ of empirical correlations of signals between pairs of
ROIs, the space $\cL_d$ is appropriate.  On the other hand, if we
chose instead to work with a thresholded version of such matrices,
then it is important that we allow for \emph{both} the presence/absence of edges by
allowing weights to be zero. 
The result of Theorem~\ref{thm:S} can be extended to include such networks, as described in the following corollary. This
leads to a manifold that possesses corners. A good introduction to
manifolds with corners can be found in standard texts on smooth manifolds
\citep[see chap.~14 in][]{Lee2006}. Moreover, this manifold is also 
a convex subset of Euclidean space.
\begin{cor}\label{cor:non-positive}
   In Theorem \ref{thm:S}, if condition (5) is replaced by 
   \begin{en}
     \item[(5${}'$)] The off-diagonal entries are non-positive, $a_{ij}\leq 0$;
   \end{en}
   then the corresponding matrix space $\cL_{d}\pri$ is a manifold
   with corners of dimension $d(d-1)/2$. Furthermore, $\cL_{d}\pri$ is a convex
   subset of an affine space in
   $\R^{d^{2}}$ of dimension $d(d-1)/2$. 
\end{cor} 

A proof of this corollary is also provided in the Supplementary Material. 
Importantly, the above theorem and its corollary indicate that the Euclidean metric
(i.e.~the Frobenius distance on the 
space of $d\times d$ matrices with real-valued entries) is a natural choice of distance
function on our spaces of Laplacians.  The metric space of interest is
therefore composed of, for example, $(\cL_{d}\pri,\rho_{F})$, where $\rho_{F}$ is the Frobenius
distance 
\begin{equation}\notag
    \rho_{F}(X,Y) := ||X-Y||_{F}^{2} = \sum_{i,j}^{d} (x_{ij} -
    y_{ij})^{2} \enskip ,
\end{equation}
for any pair of matrices $X,Y\in\cL_{d}\pri$.  As we shall see
momentarily below, in Section~\ref{sec:inference},
the concept of a Fr{\'e}chet mean and its sample-based analogue, 
as detailed in equations (\ref{eq:Frechet.mean}) and (\ref{eq:sample.Frechet.mean}),
may now be brought to bear, yielding a well-defined sense of an average of networks.

\subsection{Extensions: Implications of constraints on network topology}
In ending this section, we note that our definition of a `space of
networks' is intentionally minimal in lacking constraints 
on the topology of the networks.  However, one of the most fundamental
results that has emerged from the past 20 years of complex network
research is the understanding that real-world networks typically
(although not exclusively) tend to possess a handful of quite
marked structural characteristics.
Examples include sparseness (i.e., number of edges scaling like the
number of vertices), heavy-tailed degree distributions, and the
presence of cohesive subgraphs (a.k.a.~communities).  See chap.~8 in
\citet{Newman2010}, for example, for details and a more comprehensive
summary. In the context of neuroimaging, it can be expected that the networks
of interest will be sparse, due to a trade-off between wiring cost and
topological complexity \citep{Bullmore2012}. 
Importantly, this fact suggests that the appropriate
differential or metric measure geometry of the `space of all networks'
-- or, more formally, the space of Laplacians corresponding to such
networks -- depends on the constraints imposed on these
networks/Laplacians. 

While a detailed study of these
implications are beyond the scope of this paper,
we illustrate them through the following theorem, which extends the previous
results to the more general case of graphs composed of different
numbers of connected components. 
In particular, we can generalize Theorem \ref{thm:S} to spaces of
Laplacians representing graphs with a fixed number of components,
$\ell$. (Recall that the rank of a Laplacian is equal to $d$ minus the
number of communities in that graph.)
\begin{thm}\label{thm:S general}
The set $\cL_\ell$ of $d\times d$ matrices $E$ satisfying
\begin{itemize}
  \item[(1${}_\ell$)] $\op{Rank}(E)=\ell$,
  \item[(2)] $E$ is symmetric,
  \item[(3)] $E$ is positive semidefinite,
  \item[(4)] The sum of the entries of each column is zero,
  \item[(5)] Each off-diagonal entry is negative;
\end{itemize}
forms a submanifold of $\R^{d^2}$ of dimension 
$d\ell - \ell(\ell+1)/2$.
\end{thm}
A proof of this theorem is in the Supplementary Material. 
Intuitively, this result is stating that the number of connected
components of the average of two graphs can be smaller than the number
of components of each graph, but it cannot be larger. That is, the
average of two graphs may decrease the number of communities, but
it cannot increase that number. Indeed, when taking the
Euclidean average of several graphs with non-negative edge weights, we
can only maintain existing edges or create new edges. 

%%%%%%%%%%%%%%%%%%%%%%%%%%%%%%%%%%%%%%%%%%%%%%%%%
%%%%%%%%%%%%%%%%%%%%%%%%%%%%%%%%%%%%%%%%%%%%%%%%%
%%%%%%%%%%%%%%%%%%%%%%%%%%%%%%%%%%%%%%%%%%%%%%%%%
\section{Statistical Inference on Samples of Networks}\label{sec:inference}
Having characterized a space of networks, it becomes
possible to construct an inferential framework for comparing one or
more samples of networks. We here describe some analogues of the classical one- and two-sample
$t$-statistics in this setting. These are obtained by first selecting a
notion of averaging and deriving a central
limit theorem for sequences of network averages, next appealing to Wald-like constructions
of test statistics, and finally, utilizing recent results on high-dimensional covariance estimation.

%%%%%%%%%%%%%%%%%%%%%%%%%%%%%%%%%%%%%%%%%%%%%%%%%
\subsection{A Central Limit Theorem}\label{sec:clt}
Let $G_1,\ldots, G_n$ denote $n$ graphs, each simple and assumed to have the same number
of vertices $d$; and let $L_1,\ldots, L_n$ be the corresponding
combinatorial Laplacians.  The $L_i$'s are assumed to be 
independent and identically distributed according to a distribution
$Q$.  In the context of neuroimaging, for example,
these might be the correlation networks from resting-state fMRI images
obtained from a group of human subjects 
matched for various demographic characteristics (e.g., age, gender)
and health status (e.g., clinical manifestation of 
a given neurodegenerative disease).

The results of the previous section tell us that an appropriate sense
of distance between pairs of networks is given by the
Euclidean distance between their corresponding Laplacians.  Combining
these results with the definition of  average in equations
(\ref{eq:Frechet.mean}) and (\ref{eq:sample.Frechet.mean}), indicates 
that a principled way in which to define the average of $n$ networks
is through elementwise averaging of the entries of their Laplacians (and hence their
adjacency matrices).  Such an average is, of course, easily
computed. However, this is not always the case when computing averages on manifolds. See,
for instance, chap.~6 in \citet{Bhatia2007} for an illustration of the
difficulties that may arise, when computing the matrix mean in the
cone of positive-definite symmetric matrices with respect to the
geodesic distance on that manifold.

In the context of the 1000 FCP database, we wish to compare networks
with respect to the sex of the subjects, and over different age groups. It is thus
necessary to compute the means in each subgroup of networks. This 
was done, for example, in Figure \ref{fig:sexage}, by constructing the Euclidean
mean of the Laplacians for each group of subjects in different age groups.
Such group-specific mean Laplacians can then be interpreted as
the mean functional connectivity in each group.

The sample Fr{\'e}chet mean $\wh{L}_{n}$ is a natural statistic upon which
to build our hypothesis tests about 
the average of networks or groups of networks.  In order to do so, we
require an understanding of the behavior of $\wh{L}_{n}$  as a random
variable.  Under broad regularity conditions, $\wh{L}_{n}\to\La$
almost surely; that is, the sample Fr\'echet mean, $\wh{L}_{n}$, is a
consistent estimator of the true mean $\La$ \citep[see][]{Ziezold1977}. In
addition, under further assumptions, we can also derive a
central limit theorem for the sample Fr\'echet mean of Laplacians,
with respect to the half-vectorization map, $\phi$.
\begin{thm}\label{thm:clt}
   If the expectation, $\La:=\E[L]$, does not lie on the boundary of 
   $\cL\pri_{d}$, and $\p[U]>0$, where $U$ is an
   open subset of $\cL\pri_{d}$ with $\La\in U$, and $\cL\pri_{d}$
   defined as Corollary \ref{cor:non-positive}; and under some
   further regularity conditions (see Supplementary Material); we obtain the
   following convergence in distribution,
   \begin{equation}\notag
      n^{1/2}(\phi(\wh{L}_{n}) - \phi(\La)) 
      \lra N(0,\Sig),
   \end{equation}
   where $\Sig:=\cov[\phi(L)]$ and $\phi(L)$ denotes the
   half-vectorization of $L$.
\end{thm}
Theorem \ref{thm:clt} assumes that the true
Fr\'echet mean does not lie on the boundary of the parameter space,
which requires that all of its off-diagonal entries are non-zero. This
potentially conflicts with the fact that neuroimaging networks are
often hypothesized to be sparse \citep{Bullmore2012}. Note, however,
that we are only requiring such entries to be non-zero in expectation. 
Thus, any positive value would suffice to ensure that the true
Fr\'echet mean is away from the boundary, although structural zeros
(i.e.~zeros in the true Fr\'echet mean) would be problematic.

A proof of this theorem and the full set of assumptions are provided
in the Supplementary Material. The argument is a specialization of a general result due to
\citet{Bhattacharya2016}. The result stated in the theorem has
fundamental significance regarding our goal
of developing analogues of classical testing strategies for the
analysis of network data objects.  It is 
an asymptotic result stating that, given a sufficient number of samples
from a population of networks, an appropriately defined notion of sample average
behaves in a classical manner: It possesses a statistical distribution that is
approximately multivariate normal, centered on the population mean
$\mu$ and with covariance $\Sigma$. Note that this population covariance is
assumed to be positive definite, even though its sample estimate needs not be
positive definite, as we will see in our examples. In such cases, we
will estimate that covariance matrix using a method due to \citet{Schafer2005}.

Theorem \ref{thm:clt} can be straightforwardly extended in order to
select specific entries in the Laplacians under scrutiny. Such
sub-matrices would correspond to certain sub-graphs of interest. 
In particular, given an orthogonal projection matrix, $\Ga$, and using the fact that such
central limit results are preserved under linearity, we obtain the
following generalization of Theorem \ref{thm:clt},
\begin{equation}\label{eq:projection}
      n^{1/2}(\Ga\phi(\wh{L}_{n}) - \Ga\phi(\La)) 
      \lra N(0,\Ga\Sig\Ga\pri).
\end{equation}
In the context of neuroimaging, this may allow to identify subnetworks
of specific interest, such as the so-called default mode network, for
instance \citep{Greicius2003}. We will study the properties of this
neuroanatomical network in the sequel. 

%%%%%%%%%%%%%%%%%%%%%%%%%%%%%%%%%%%%%%%%%%%%%%%%%
\subsection{One-sample, Two-sample and $k$-sample Tests}\label{sec:test}
As an immediate consequence of this central limit theorem, we can
define natural analogues of classical one- and $k$-sample hypothesis
tests. Consider, for example, the null hypothesis 
that the expectation $\Lambda=\E[L]$ is equal to some pre-specified
value, i.e., $H_0: \Lambda = \Lambda_0$.
In the context of neuroimaging, the choice of $\Lambda_0$ might
correspond to a reference connectivity pattern, derived from a large
study, such as the 1000 FCP, for instance.
In addition to the conditions
stated in Theorem \ref{thm:clt}, let us now assume that the
true covariance matrix, $\Sig$, is \textit{non-singular}. 
Moreover, it is also assumed that the target Laplacian, $\La_{0}$, is
known. 
\begin{cor}\label{cor:one-sample}
Under the assumptions of Theorem \ref{thm:clt}, and under the null
hypothesis $H_0: \E[L]=\La_{0}$, the test statistic,
\begin{equation}\notag
   T_{1}:=n\big(\phi(\wh{L}) -\phi(\La_{0})\big)\pri
   \wh\Sig^{-1}\big(\phi(\wh{L}) - \phi(\La_{0})\big),
\end{equation}
converges to a $\chi^{2}$-distribution with $p:=\binom{d}{2}$ degrees of
freedom, and where
$\wh{\Sig}:=1/(n-1)\sum_{i=1}^{n}(\phi(L_{i})-\phi(\wh{L}))(\phi(L_{i})-\phi(\wh{L}))\pri$
denotes the sample covariance.
\end{cor}

See Theorem 5.2.3 of \citet{Anderson2003}, for a proof.  
Similarly, one can also construct a statistical test for two independent
samples using the same framework. Assume that we have two
independent sets of Laplacians of dimension $d\times d$, and consider the problem of
testing whether these sets have in fact been drawn from the same
population. Each sample of Laplacians has the
form, $L_{in_{j}}$, where $i=1,\ldots,n_{j}$; for every
$j=1,2$. The population means are
denoted $\La_{j}$, while the sample means of these sets of
Laplacians are denoted by $\wh{L}_{j}$. Then, as a direct corollary to
Theorem \ref{thm:clt}, we also have the following asymptotic result. 
\begin{cor}\label{cor:two-sample}
   Assume that every $\La_{j}$ does not lie on the boundary of
   $\cL\pri_{d}$, and that $\p[U]>0$, where $U$ is an open
   subset of $\cL\pri_{d}$, such that $L_{j}\in U$, for each
   $j=1,2$. Moreover, also assume that $n_{j}/n\to p_{j}$ for
   every sample, with $n:=\sum^{2}_{j=1}n_{j}$, and $0<p_{j}<1$. Then, under
   the null, $H_{0}:\La_{1}=\La_{2}$, we have
   \begin{equation}\notag
     T_{2}:=  \frac{n_{1}n_{2}}{n_{1}+n_{2}}(\phi(\wh{L}_{1})-\phi(\wh{L}_{2}))\pri
     \wh{\Sig}^{-1}(\phi(\wh{L}_{1})-\phi(\wh{L}_{2})) 
     \longrightarrow \chi^{2}_{p},
   \end{equation}
   where $\wh{L}_{j}$ denotes the $j\tth$ sample mean,
   and $\wh{\Sig} := (\sum_{j=1}^{2}n_{j}\wh{\Sig}_{j})/(n_{1}+n_{2}-2)$ is the
   pooled covariance estimate, with the $\wh{\Sig}_{j}$'s denoting
   the individual covariance matrices of each subsample.
\end{cor}
Finally, we can also derive a test statistic $T_k$, to test 
that $H_{0}:\La_{1}=\ldots=\La_{k}$ is true. This can be performed using an
analogue of Wilks's statistic, by partitioning the variance of interest,
using the asymptotic result in Theorem \ref{thm:clt}.

%%%%%%%%%%%%%%%%%%%%%%%%%%%%%%%%%%%%%%%%%%%%%%%%%
\subsection{Covariance Estimation}\label{sec:covariance estimation}
We note that in order to use any of the above results in a practical setting,
we must have knowledge of the covariance matrix $\Sigma =
\cov[\phi(L)]$.  It can be
expected that we must use a sample-based estimate.  However, because
the dimension of this matrix is $O(d^2)\times O(d^2)$, and
the sample size $n$ is potentially much smaller than $O(d^2)$, the traditional sample
covariance $\wh\Sigma$ is likely to be numerically unstable, and is not
guaranteed to be positive definite.  

Fortunately, the development of estimators of $\Sigma$ 
in such low-sample/high-dimension contexts has been an 
active area of statistical research over the past few years.
Typically, borrowing regularization strategies from the field of nonparametric function
estimation, optimization of a cost function combining  Frobenius norm
or penalized maximum likelihood
with a regularization term yields a convex optimization problem that
can be solved efficiently. Generally, the
choice of a regularization term is linked to the assumed structure of the
covariance matrix -- for example, assumptions
of banding \citep{Bickel2008} or sparseness
\citep{Bickel2008a,Cai2011a,Karoui2008}.
There is also a substantial recent literature on the closely related
problem of estimating the inverse covariance matrix
$\Sigma^{-1}$. See \citet{Cai2011} for a recent example and
associated citations. 
 
In our context, there is little understanding of how the covariance
matrices of the off-diagonal entries of graph Laplacians should
behave. Accordingly, as an alternative to the sample covariance, we have adopted
a shrinkage estimator due to \citet{Schafer2005}, which is
particularly well-suited to large data sets. The method of
\citet{Schafer2005} is a popular generic method for
high-dimensional covariance estimation, which works through shrinkage
towards substructures. This was deemed sufficiently flexible for our
purpose.

Moreover, since in finite samples, the estimator $\wh{\Sig}$ may not necessarily be a
positive definite matrix; we have therefore adopted
an algorithm due to \citet{Higham2002}, in order to locate the nearest
positive definite matrix in Frobenius norm \citep[see
also][]{Cheng1998}. The resulting
matrix, say $\widetilde{\Sig}_{PD}$, is then used
in place of $\widehat\Sigma$ in the test statistics described in the
previous section.

%%%%%%%%%%%%%%%%%%%%%%%%%%%%%%%%%%%%%%%%%%%%%%%%%
\subsection{Visualization of Differences}\label{sec:visualization}
The contribution of each node to the test statistics in the one- and
two-sample tests can be visualized through a linear decomposition
of the above test statistics. Consider, for example, the one-sample
statistic, $T_{1}$. By taking the square-root of the inverted
covariance matrix, this particular quantity admits the following
decomposition, $T_{1}/n=\wh\la\pri\wh\la$, in which
$\wh\la:=\Sig^{-1/2}(\phi(\wh{L})-\phi(\La_{0}))$. Consequently,
we can reformulate this one-sample test as a sum of squares, $T_{1} =
n\sum_{i=1}^{p}\wh\la_{i}^{2}$, with
$p:=\binom{d}{2}$. The square of the $i\tth$ entry of $\wh\la$,
can then serve as an indicator of the contribution of the $i\tth$
edge to the value of $T_{1}$. 

Similarly, this visualization strategy can be extended to
the comparison of two groups. That is, the statistic $T_{2}$ in
Corollary \ref{cor:two-sample} can be represented as
$(n_{1}+n_{2})/(n_{1}n_{2})T_{2} = \sum_{j=1}^{2}\wh\la_{j}\pri\wh\la_{j}$,
where the two $p$-dimensional vectors, $\wh\la_{1}$ and $\wh\la_{2}$,
are defined as $\wh\la_{j}:=\wh{\Sig}^{-1/2}(\phi(\wh{L}_{j})-\phi(\wh{L}))$ for
$j=1,2$. Consequently, we can decompose the weighted version of
$T_{2}$ as a sum of $p$ terms of the form, 
\begin{equation}\notag
    \frac{n_{1}+n_{2}}{n_{1}n_{2}}T_{2} = \sum_{j=1}^{2}\wh\la_{j}\pri\wh\la_{j}
     = \sum_{j=1}^{2}\sum_{i=1}^{p}\wh\la^{2}_{ij} 
     = \sum_{i=1}^{p}\bigg(\sum_{j=1}^{2}\wh\la^{2}_{ij}\bigg)
     =: \sum_{i=1}^{p}\wh\ka_{i};
\end{equation}
in which $\wh\la_{ij}$ stands for the $i\tth$ element of
$\wh\la_{j}$, and $\wh\ka_{i}:=\sum_{j=1}^{2}\wh\la^{2}_{ij}$. Therefore,
as for $T_{1}$, each $\wh\ka_{i}$ can be treated as the specific
contribution of the $i\tth$ edge to the value of $T_{2}$. 

In the sequel, we will plot the $\wh\ka_{i}$'s in order to provide a fine-grained
visualization of the differences between the families of networks
under scrutiny. The empirical distribution of the edgewise contributions,
$\wh\ka_{i}$'s, to the estimate of the statistic
$T_{2}$, will be computed; and we will report all values
above a certain threshold. In Figure
\ref{fig:mass}, for instance, we have plotted the edgewise
contributions above the $85\tth$ and $95\tth$ percentiles of the
distribution of the $\ka_{i}$'s. Since each $\wh\ka_{i}$ is positive,
our strategy consists in partitioning $T_{2}$ into a sum of positive
values; and the plotted $\wh\ka_{i}$'s in Figure \ref{fig:mass} represent the
largest such values.

%%%%%%%%%%%%%%%%%%%%%%%%%%%%%%%%%%%%%%%%%%%%%%%%%
%%%%%%%%%%%%%%%%%%%%%%%%%%%%%%%%%%%%%%%%%%%%%%%%%
%%%%%%%%%%%%%%%%%%%%%%%%%%%%%%%%%%%%%%%%%%%%%%%%%
\section{Simulation Studies}\label{sec:sims}
In this empirical study, we evaluate the statistical power of the
two-sample test $T_2$ for Laplacians, under different choices of number of vertices and
for increasing sample sizes. We simulate network-based data
for $n$ subjects in each group, and focus our attention on
two-sample experimental designs. Motivated by the neuroimaging application
underlying the methodological development just described, 
the data generating process relies on
(i) the selection of a network topology and the construction of an
associated covariance matrix, (ii) the generation of multivariate
time series for each network model, and (iii) the construction of 
subject-specific Laplacians based on the covariance matrices.

%%%%%%%%%%%%%%%%%%%%%%%%%%%%%%%%%%%%%%%%%%%%%%%%%
\subsection{Network Topologies}\label{sec:topology}
In these simulations, we consider two types of network topology,
specified through binary matrices, $A_{1}$ and $A_{2}$, of order $d\times d$. 
Firstly, we consider a block-diagonal structure $A_{1}$, which
represents the grouping of several vertices into two homogeneous
communities, 
\begin{equation}\notag
    A_{1} :=
    \begin{pmatrix}
        X & R\\
        R & Y
    \end{pmatrix},
\end{equation}
where $X$ and $Y$ are square matrices of dimensions $\lceil d/2
\rceil$ and $\lfloor d/2 \rfloor$, respectively. The elements of $X$
and $Y$ are given a value of 1 according
to independent Bernoulli variates with proportion $p_{1}:=4/d$; whereas
the elements of $R$ take a value of 1 with a probability of
$p_{2}:=1/(2d)$. These choices of $p_{1}$ and $p_{2}$ ensure that the
corresponding block models are \textit{sparse} in the sense that their
numbers of edges are proportional to their numbers of vertices, as $d$ grows.

\begin{figure}[t]
   \scriptsize 
   \centering   
   %%% Two simulated structures
   \textbf{(A)} \textit{Block Diagonal}
   \hspace{1cm}\textbf{(B)} \textit{Small-world}\\ 
   \vspace{.25cm}
   \includegraphics[width=2.5cm]{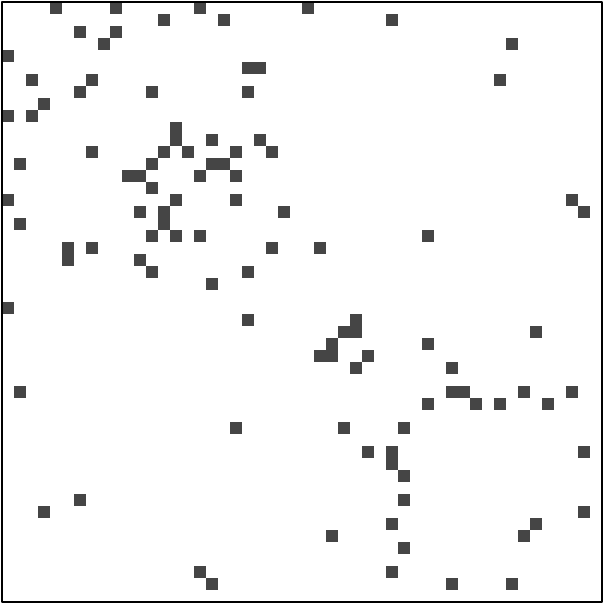}
   \hspace{1cm}
   \includegraphics[width=2.5cm]{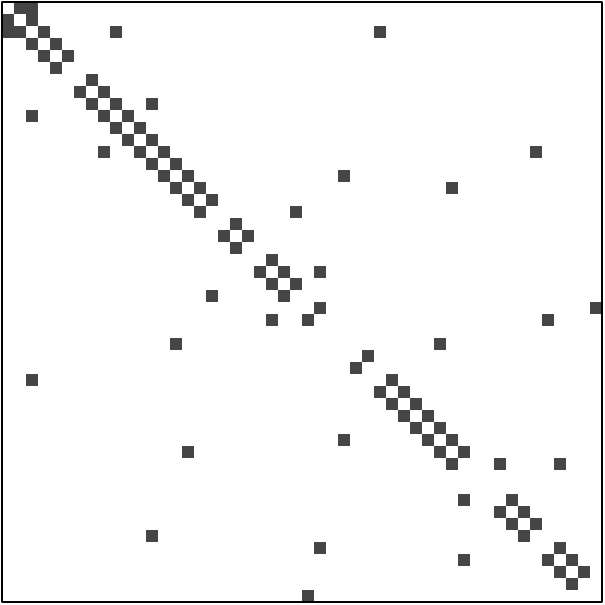}\\
   \caption{Simulated matrices over $d=50$ vertices. In panels (A) and
     (B), matrices with a block-diagonal structure and a small-world topology
     are respectively represented. 
     \label{fig:networks}}
\end{figure}

Secondly, we specify a small-world network structure, $A_{2}$,
by constructing a regular network with a ring topology, whose number of edges is
taken to be proportional to $d$, which again enforces sparsity.
The edges of this network are then randomly rewired
\citep{Watts1998}. The choice of $N_{e}$ --the number of edges-- is here motivated by a desire
to maintain some level of comparison between the block-diagonal model and the
small-world topology. Using such $N_{e}$'s, we ensure that both types of
networks have approximately the same number of edges.
These two families of network topologies are illustrated in Figure
\ref{fig:networks} for simulated networks of size $d=50$.

For both of these models, we generated mean covariance matrices,
$S_{gm}$'s, where $g=1,2$ denoting the group of subjects,
and $m=1,2$ denoting the block model and small-world model,
respectively. These were constructed using a mixture model, based on the
binary matrices, $A_{m}$'s. The $S_{2m}$'s were expressed as a
function of the $S_{1m}$'s. For the diagonal elements of the $S_{1m}$'s,
\begin{equation}\notag
    S_{aa,1m} \stack{\iid}{\sim} \op{exp}(\la), \qquad a=1,\ldots,d;
\end{equation}
whereas the off-diagonal elements of the $S_{1m}$'s are constrained
by the corresponding off-diagonal elements in the adjacency matrices,
$A_{m}$'s, as follows,
\begin{equation}\notag
    S_{ab,1m}|\te{A}_{ab,m} \stack{\ind}{\sim} 
    |\te{A}_{ab,m}N(\mu_{1},\sig^{2}) + (1-\te{A}_{ab,m})N(\mu_{2},\sig^{2})|; 
\end{equation}
for every $a\neq b$, and where the parameters of the mixture model are given
the following values, $\la:=4$, $\mu_{1}=1$, $\mu_{2}=0$ and $\sig^{2}=.2$ for
all simulation scenarios; thereby producing a high signal-to-noise
ratio, permitting to distinguish between the different types of
entries in the matrices, $S_{1m}$'s.
Note that none of the simulation scenarios guarantees that
the resulting $S_{1m}$'s are positive definite. Consequently, we 
projected the resulting matrices to the nearest positive definite
matrices in the Frobenius norm, using the method described in Section
\ref{sec:covariance estimation}. Once the $S_{1m}$'s were obtained,
they were fixed for each scenario, and used to generate the covariance
matrix in the second group as follows, $S_{2m}:= C(\eta-1)S_{1m}$,
where $\eta$ controlled the distance between the two population means,
which was interpreted as the effect size; and the constant $C$ was set to a
small value, $C:=0.03$, throughout the simulations.
%%%%%%%%%%%%%%%%%%%%%%
%%%%%%%%%%%%%%%%%%%%%%

%%%%%%%%%%%%%%%%%%%%%%%%%%%%%%%%%%%%%%%%%%%%%%%%%
\subsection{Noise Models}\label{sec:noise}
Resting-state or default-mode brain networks have been investigated by
a large number of researchers in neuroimaging
\citep{Thirion2006,Beckmann2005a}. The main difficulty in simulating
these networks stems from the absence of a prior to produce such
resting-state patterns of activities \citep{Leon2013,Kang2012}. 
For each subject, we here constructed a set of $d$ sequences
of $T$ realizations, where $d$ represents the number of ROIs, and $T$
denotes the total number of time points. These sequences of
realizations were drawn from a multivariate Gaussian, such that for
every subject, $i=1,\ldots,n$, the random vectors,
$X_{itgm}\in\R^{d}$, were given by
\begin{equation}\notag
    X_{itgm} \stack{\iid}{\sim} N_{d}(0,S_{gm}),
    \qquad\forall\;t=1,\ldots,T;
\end{equation}
where $g=1,2$ denotes group affiliation, and $m=1,2$ denotes the
choice of underlying adjacency matrix: block-diagonal model and
small-world model.
%%%%%%%%%%%%%%%%%%%%%%
\begin{figure}[t]
   \includegraphics[width=13cm]{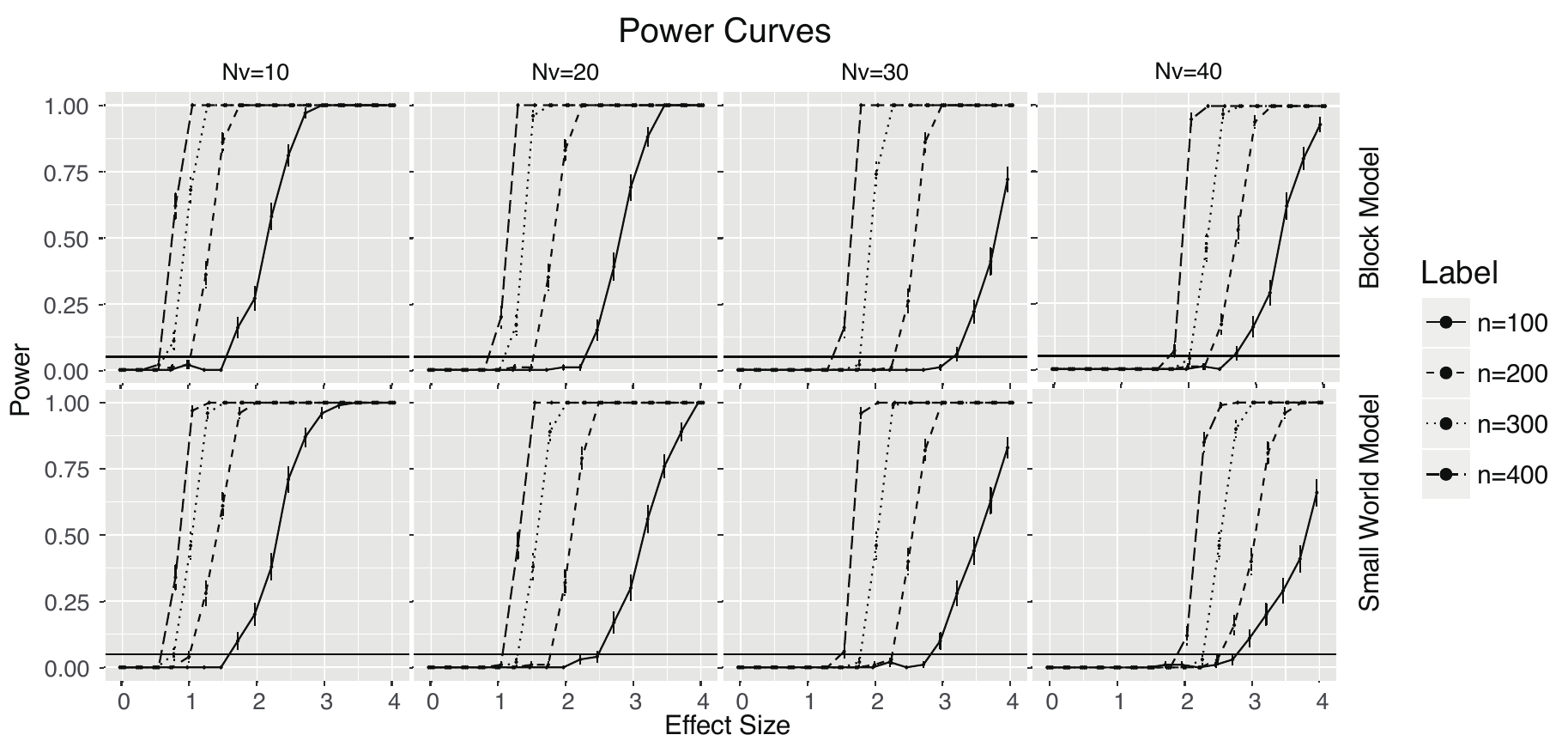}
   \caption{Power curves for the simulated two-sample tests
     using the \textit{covariance} estimation procedure, under a
     multivariate Gaussian model, with error bars based on one 
     standard error from the mean. 
     The $y$-axis indicates the probability of rejecting
     the null hypothesis when it is false; whereas the $x$-axis
     is a proxy measure of effect size (see Section \ref{sec:sims result}).
     These results are presented for networks on $d=10,20,30$ and $40$
     vertices, with different sample sizes and over $T=50$
     time points, and based on $100$ iterations per
     condition with respect to the block (top row) and small-world
     (bottom row) topologies. A horizontal line has been added to
     indicate a power of $0.05$.\label{fig:sim_cov}}
\end{figure}
%%%%%%%%%%%%%%%%%%%%%%

%%%%%%%%%%%%%%%%%%%%%%%%%%%%%%%%%%%%%%%%%%%%
\subsection{Simulation Results}\label{sec:sims result}
Four main factors were made to vary in this set of simulations.
In line with the subsequent real-data analysis, we considered sample
sizes of $n=100,200,300$ and $400$ per group. This was deemed representative of the
number of subjects found in most neuroimaging studies. 
Secondly, we varied network sizes, with $d$ taking
values $10,20,30$, and $40$. 
This range of network sizes allowed us to identify
the effect of network size on the statistical power of our test.
Larger dimensions were expected to decrease power. 

In each of these scenarios, we computed the statistical power of the
two-sample tests, using different effect sizes. Here, the effect size
was defined with respect to the value of the parameter $\eta$. Recall
that $\eta$ controlled the distance between the two population means,
such that $S_{2m}:= C(\eta-1)S_{1m}$.
For each set of conditions, the simulations were repeated 100 times
in order to obtain an empirical estimate of the theoretical power of
the two-sample test statistic for Laplacians, under these conditions.

The results of these simulations are reported in Figure
\ref{fig:sim_cov}. The power of the two-sample
test for Laplacians was found to be empirically well-behaved, 
for all the scenarios considered. In particular, this was true for
both the block-diagonal and small-world topologies, as illustrated in
the first and second row in Figure \ref{fig:sim_cov}. As
expected, the power of the test tended to increase with larger sample
sizes, albeit that increase was mitigated by the size of the
underlying networks. 

%%%%%%%%%%%%%%%%%%%%%%%%%%%%%%%%%%%%%%%%%%%%%%%%%
%%%%%%%%%%%%%%%%%%%%%%%%%%%%%%%%%%%%%%%%%%%%%%%%%
%%%%%%%%%%%%%%%%%%%%%%%%%%%%%%%%%%%%%%%%%%%%%%%%%
\section{Analysis of the 1000 FCP Data Set}\label{sec:data}
Different aspects of the 1000 FCP data set were considered. Firstly,
we used a one-sample test for comparing the Laplacian mean to a
subsample of the data. We then tested for sex and age differences using
the two- and $k$-sample tests for Laplacians. Finally, we analyzed
differences in subnetworks, including the default-mode network (DMN).
After excluding subjects for which demographics data were incomplete,
we obtained a sample size of $n=1017$. 

%%%%%%%%%%%%%%%%%%%%%%
\subsection{Inference on Full Data Set}
%%% Reference Connectome
As described in Section \ref{sec:bg.neuro}, the 1000 FCP data provides
a unique opportunity for neuroscientists to extract a reference
template of human connectivity. We tested the
reliability of that template using a one-sample Laplacian test for
some random subsample of the data. We computed the reference mean
Laplacian over the full FCP sample, which is here treated as a
\textit{population parameter}, $\La_{0}$. This was compared with a
large random subsample of $917$ subjects --that is, after removing 100
subjects from the original FCP data. We then tested for the null
hypothesis that the sample mean, $\wh{L}_{1}$, was equal to the
reference mean $\La_{0}$. As expected, the test failed to reject the
null hypothesis $(T_{1}=926, \df=\binom{50}{2}, p\approx1)$, since the
sample and reference means were drawn from the same population.

\begin{figure}[t]
   \raggedright
   %%% Two simulated structures:
   % \textbf{(A) Effect of Sex} \\
   \vspace{.1cm}
   \hspace{.6cm} (A) Mass-univariate analysis 
   \hspace{2.45cm} (B) Multivariate analysis \\
   \vspace{.1cm}
   \hspace{.35cm}\textit{Uncorrected}\hspace{1.50cm}\textit{Corrected}
   \hspace{1.4cm}\textit{$\geq85\tth$ Percentile}\hspace{.65cm}\textit{$\geq95\tth$ Percentile}\\
   \vspace{.1cm}
   \includegraphics[width=2.5cm]{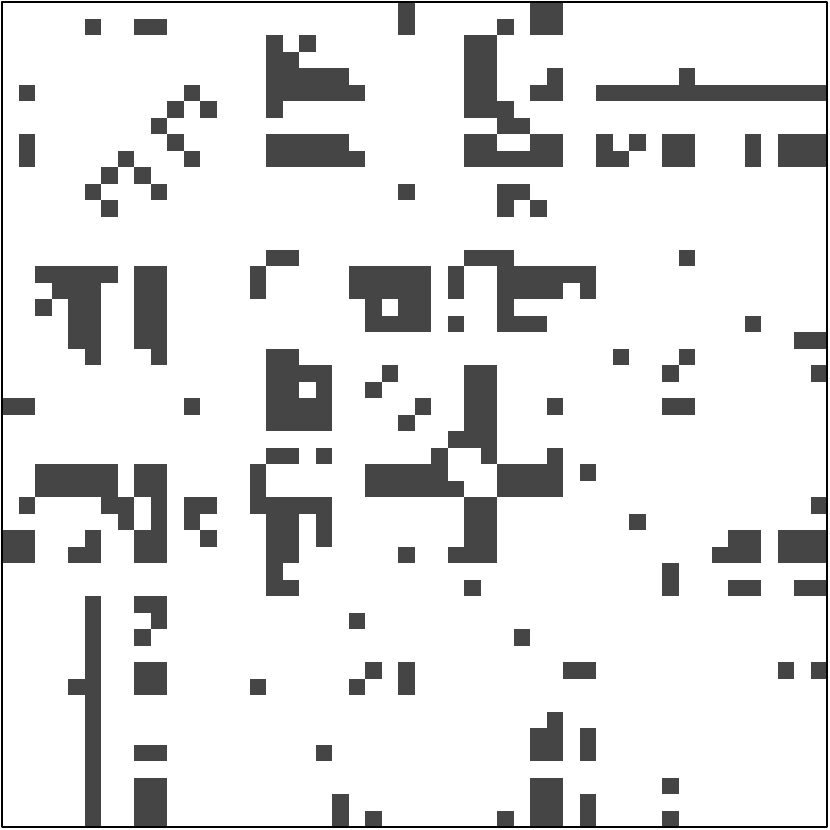}
   \hspace{.25cm}
   \includegraphics[width=2.5cm]{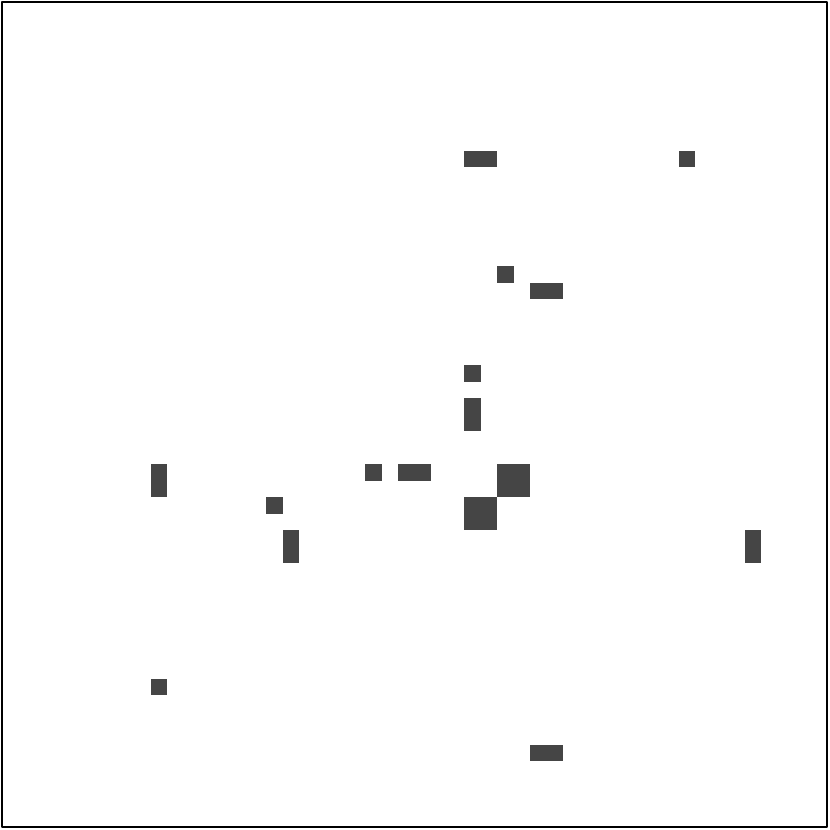}
   \hspace{.75cm}
   \includegraphics[width=2.5cm]{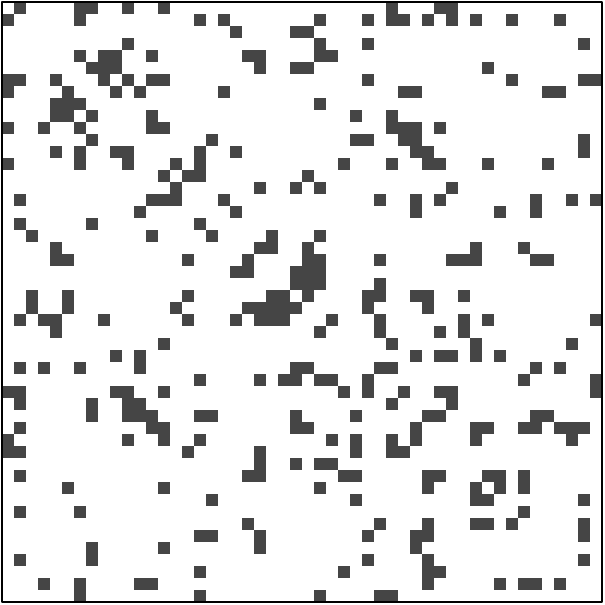}
   \hspace{.25cm}
   \includegraphics[width=2.5cm]{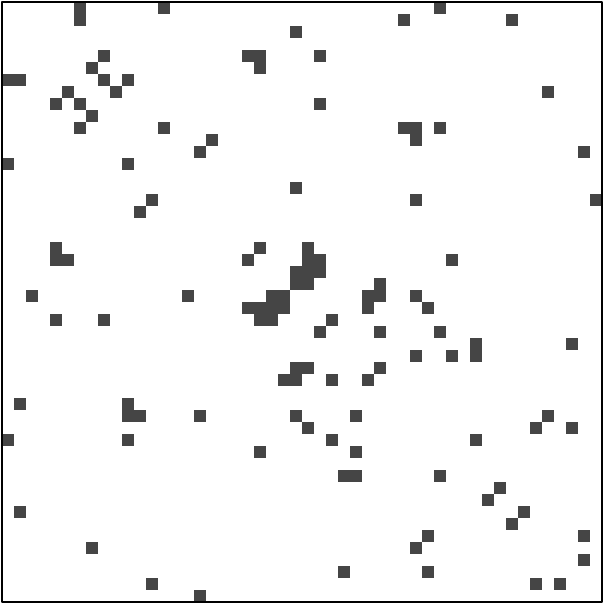}
   \caption{Comparison of edgewise differences using either a standard
     mass-univariate analysis, or our proposed multivariate method.
     In panel (A), mass-univariate analyses were conducted to test for local
     differences in connectivity due to sex in the full FCP
     data set ($n=1,017$). In each case, $\binom{50}{2}$ tests were
     performed independently for each of the off-diagonal entries in
     the Laplacians. The first matrix denotes the entries that were
     found to be significantly different between the groups at
     $\al=.05$; whereas the second matrix represents the significant
     entries after Bonferroni correction. Black denotes significant
     entries. In panel (B), the same edgewise effects are
     reported for our multivariate approach, using the visualization
     method described in Section \ref{sec:visualization}. The edges
     whose contribution to the $T_{2}$-statistic is greater
     than the $85\tth$ and $95\tth$ percentiles (of the distribution
     of such individual contributions) have been plotted in black in the
     third and fourth panels, respectively.
     \label{fig:mass}}
\end{figure}

%%% Sex Differences.
The partitioning of the 1000 FCP data set by sex is provided in Figure
\ref{fig:sexage}(A). This consists of $n_{f}=555$ female and
$n_{m}=462$ male subjects. We tested
whether such sex differences were significant using the two-sample
test for Laplacians. The null hypothesis of no group differences was
rejected with high probability ($T_{2}=1,689.5, \df=\binom{50}{2},
p<0.001$). These results should be compared with the use of a mass-univariate
approach, in which a single hypothesis test is run for each voxel. The
significant voxel-level differences detected using a mass-univariate
approach for sex, is reported in Figure \ref{fig:mass}. 

%%% Quantized Age.
Subjects in the 1000 FCP database can also be grouped according to
age. In Figure \ref{fig:sexage}(B), we have divided the
FCP sample into three subgroups of approximately equal sizes, with 
386, 297, and 334 subjects; for subjects younger than 22,
between 22 and 32, and older than 32, respectively. The $k$-sample
Laplacian test (or Wilks's Lambda) was performed to evaluate the hypothesis stating that
these $k=3$ groups were drawn from the same population. The
null hypothesis was also rejected with high probability in this case
($\La=0.106, n=1,017, \df=\binom{40}{2}, p<0.001$). (For computational
convenience, we here restricted our attention to networks with 40
nodes, which yielded invertible sample covariance matrices for the
Wilks's test.) 

%%%%%%%%%%%%%%%%%%%%%%
%%%%%%%%%%%%%%%%%%%%%%
%%%%%%%%%%%%%%%%%%%%%%
\subsection{Inference on Partial Data Set}
The results of the previous section were compared with another
analysis based on a small subset of connectomes. The 1000 FCP data set
is indeed exceptionally large for the field of neuroimaging. By contrast,
most papers using MRI data tend to report results based on smaller
data sets, commonly containing between 20 and 100 subjects. Here, we
have replicated the various statistical tests described in the last section
for such small sample sizes, in order to produce an analysis more reflective
of what might be performed by, say, a single lab. 

The conclusions of the network-level tests for the different
hypotheses of interest were found to be robust to a large decrease in
sample size. As for the full data set, sex differences remain
close to significance ($T_{2}=836.76, \df=\binom{40}{2},
p=0.077$), when solely considering 100 female and 100 male
subjects. Note, however, that our proposed global test failed to reject the null
hypothesis when considering smaller data sets. Indeed, we
restricted our attention to smaller subsets of subjects, composed of
20 cases in each group, and such a test did not reject the null hypothesis
($T_{2}=514.96, \df=\binom{40}{2}, p\approx 1$). 

These results should be contrasted with the use of a
mass-univariate approach. We compared the
conclusions of a network-level Laplacian test for sex, with the ones of a 
mass-univariate approach based on 100 female and 100 male subjects. 
No local differences were here found, after correcting for
multiple comparisons, and solely one edge out of $\binom{40}{2}$ was
found to significantly differ between groups at a threshold of $0.001$. 
This highlights one of the important advantages
of using a global test in this context. While the mass-univariate
approach fails to detect any sex differences at the local level, our
proposed global test, by contrast, had sufficient power to reject 
the null hypothesis at a global level.
%%%%%%%%%%%%%%%%%%%%%%
\begin{figure}[t]
   \raggedright
   \hspace{.8cm}\textbf{(A) Sex (DMN)} \hspace{3.75cm} \textbf{(B) Age (DMN)}\\
   \vspace{.25cm}
   \hspace{0.75cm}\textit{Female} \hspace{1cm} \textit{Male}
   \hspace{2.50cm}$x\leq 22$ \hspace{.6cm} $22< x \leq 32$ \hspace{.52cm} $32<x$\\
   \vspace{0.10cm}
   %%% Sex laplacians.
   \hspace{0.4cm}
   \includegraphics[width=1.5cm]{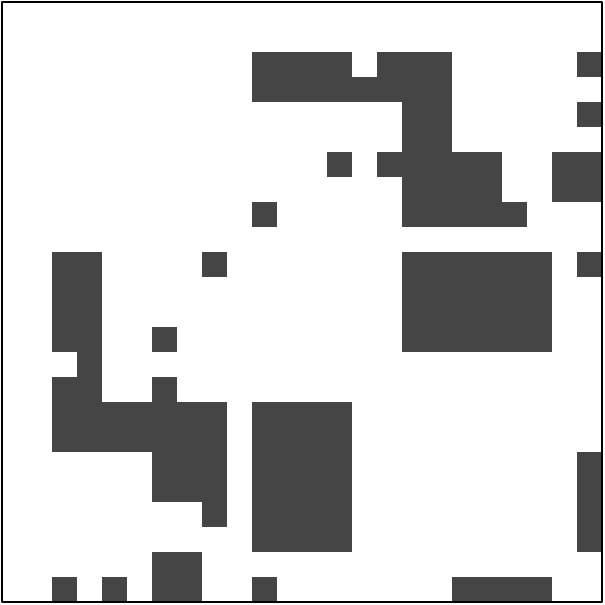}
   \hspace{.25cm}
   \includegraphics[width=1.5cm]{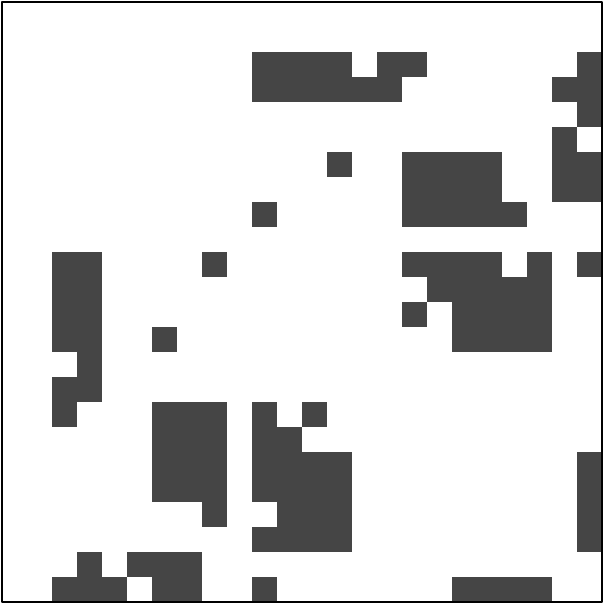}
   %%% Age Laplacians.
   \hspace{1.90cm}
   \includegraphics[width=1.5cm]{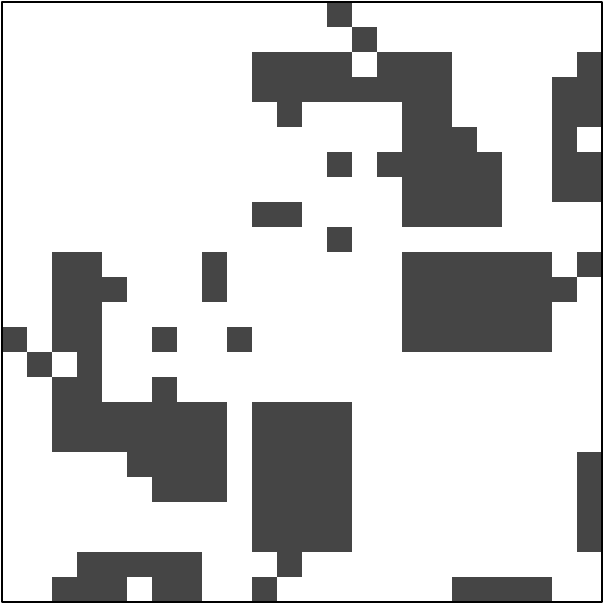}
   \hspace{.25cm}
   \includegraphics[width=1.5cm]{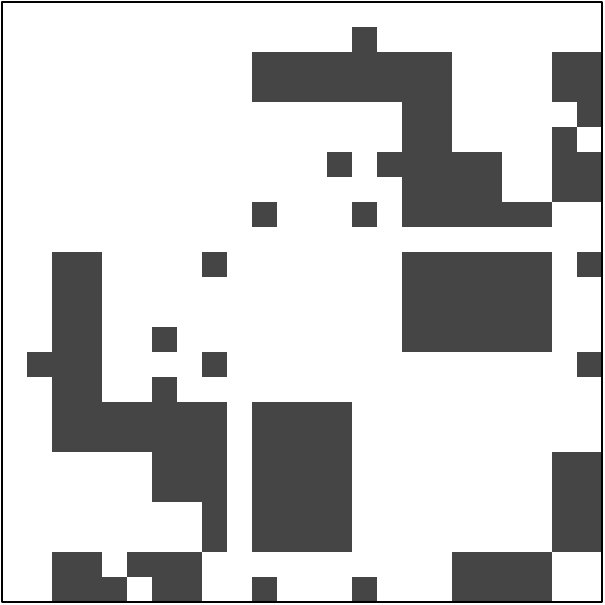}
   \hspace{.25cm}
   \includegraphics[width=1.5cm]{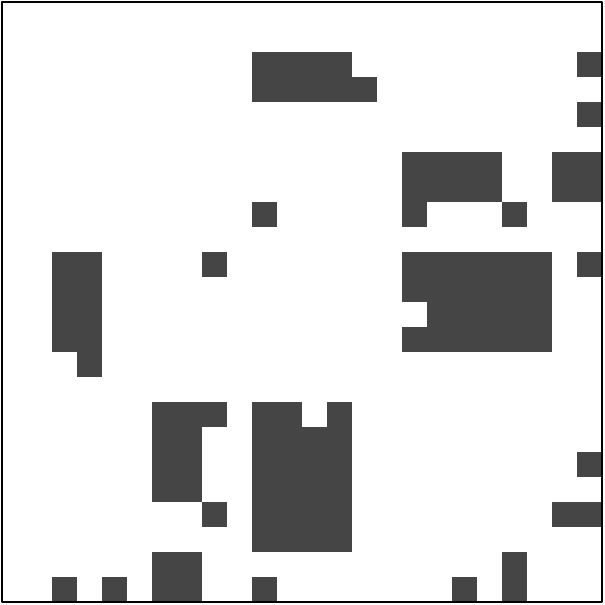}
   \caption{Descriptive statistics for Default-Mode Network (DMN) in
     the 1000 FCP data set. In panels (A) and (B), we have provided the mean
     Laplacians after dividing the sample according to sex and age,
     respectively. As before, the
     Laplacians have been binarized with respect to the $75\tth$
     percentile in the distribution of the FCP database. 
     \label{fig:dmn}}
\end{figure}
%%%%%%%%%%%%%%%%%%%%%%

%%%%%%%%%%%%%%%%%%%%%%%%%%%%%%%%%%%%
\subsection{Default-Mode Network}\label{sec:default-mode network}
The Default-Mode Network (DMN) is a widely studied portion of the
functional network characterizing brain activity in both humans and
animals \citep{Greicius2003,Buckner2008}. This network tends to be
active, when an individual is not engaged in a cognitive task. 
The DMN is composed of a set of hubs that include the precuneus, posterior
cingulate, medial prefrontal cortex, and angular gyri, as well as
prefrontal cortices, temporo-parietal junctions, the hippocampi,
and the parahippocampi. In the parcellation template used in this
paper, these regions corresponded to $d=24$ AAL areas. 

We tested for the effect of sex in the
full FCP sample by applying the projection method described in Equation
\eqref{eq:projection}. The hypothesis of no difference between males
and females was not rejected for the DMN network ($T_{2}=1,128,
\df=\binom{24}{2}, p\approx1$). (The mean Laplacians for these subnetworks
are reported in Figure \ref{fig:dmn}.) This demonstrates that such
multivariate methods also tends to lose power, when restricted to
subnetworks. 

%%%%%%%%%%%%%%%%%%%%%%%%%%%%%%%%%%%%%%%%%%%%%%%%%
%%%%%%%%%%%%%%%%%%%%%%%%%%%%%%%%%%%%%%%%%%%%%%%%%
%%%%%%%%%%%%%%%%%%%%%%%%%%%%%%%%%%%%%%%%%%%%%%%%%
\section{Discussion}\label{sec:discussion}
In this paper, we have analyzed a large neuroimaging data set, using a novel
framework for network-based statistical testing.  The development of
this framework is grounded in
a formal asymptotic theory for network averages, developed
within the context of a well-defined notion of the space of graph Laplacians.
Importantly, we have showed that using the global
tests that result from our framework
may provide the researcher with decidedly more statistical power than when using
a mass-univariate approach, which is the standard approach in the field. 

To the best of our knowledge, we are the first to ascribe a notion of a `space' to the
collection of graph Laplacians and to describe the geometrical
properties of this space.  While we have found
it convenient for the purposes of exposition simply to summarize these
results in the main body of the paper, and to collect details 
in the appendices, it is important to note that this initial step is
crucial in allowing us to bring
to bear recent probabilistic developments in the field of shape analysis to produce our key central limit theorem,
upon which the distribution theory for our tests lies.  We note too that the framework we offer is quite general
and should, therefore, as a result be quite broadly applicable.  Nevertheless, this initial work also 
has various limitations, and furthermore sets the stage for numerous directions for extensions, which we 
describe briefly below.

%%%%%%%%%%%%%%%%%%%%%%%%%%%%%%%%%%%%
\subsection{Limitations}\label{sec:limitations}
It can be expected that there be a tradeoff in the performance of our tests
between sample size $n$ and the dimension $d$ of the networks in the sample.
This expectation is confirmed in our simulations, where one can
observe that for a given sample size $n$, the rate of type I error
increases beyond the nominal rate, as $d$
increases. Since our test can be seen to be equivalent to a Hotelling $T^{2}$ on the
off-diagonal elements of the Laplacians, it follows that sample sizes of 
order $O(d^{2})$ would be required to control for this increase in type I error
rate. For the analysis of the full FCP data set, this condition was approximately satisfied,
since this data set contains more than 1000 subjects, and we were
here comparing networks with $50$ vertices. 
In their current forms, such global statistical
tests may therefore be most applicable to very large data sets, or
to relatively small networks.  However, our analysis of the smaller subsets of the FCP data
(i.e., mimicking analysis at the level of a single lab) suggests that even at low sample 
sizes the test is well-powered against the alternative of differences
in network group averages.

Computationally, the method employed in this paper was also challenging
since the application of the Laplacian test required the inversion of
a large covariance matrix. We have here resorted to different methods
to facilitate this process including the use of modern shrinkage estimation
techniques \citep{Schafer2005}, as well as the modification of the resulting
sample covariance matrix estimates in order to force positive definiteness
\citep{Cheng1998,Higham2002}. Practically, however, such methods
remain computational expensive, and may therefore limit the
size of the networks that one may wish to consider when using such
Laplacian tests. 

Finally, observe that the networks of interest in this paper have
been constructed using fMRI data. This preliminary step
necessitated the estimation of covariance matrices for each
subject, and such estimation has not been directly taken into
account in the final analysis. Further research may
therefore need to adopt a global modelling strategy in which the
uncertainty at the first level of the analysis is propagated to the
second level, in which we compare groups of subject-specific
networks.

%%%%%%%%%%%%%%%%%%%%%%%%%%%%%%%%%%%%
\subsection{Extensions}\label{sec:extensions}
In our work here (specifically, as described in Section~\ref{sec:char.nets}) we show
that the `space' of networks -- \emph{without any structural
constraints} -- behaves `nicely' from the mathematical perspective,
and therefore we are able to develop a corresponding
probability theory and statistical methods for one- and two-sample
assessment of network data objects.  However, one of the most
fundamental results that has emerged from the past 20 years of complex
network research is the understanding that real-world networks
typically (although not exclusively) in fact tend to possess a handful
of quite marked structural characteristics. For example, most networks
are relatively sparse, in the sense that the number of edges is on the
same order of magnitude as the number of vertices.  Other common key
properties include heterogeneous degree
distributions, cohesive subgraphs (a.k.a.~communities), and small-world
behavior \citep[see][chap.8]{Newman2010}.

The ubiquity of such characteristics in real-world networks has been
well-established.  Importantly, this fact suggests that the
appropriate (differential or metric measure) geometry of the `space of
all networks' -- or, more formally, the space of Laplacians
corresponding to such networks -- depends both
on the constraints imposed on these networks/Laplacians and the
geometry chosen for the larger space ${\mathcal P}$ of PSD
matrices. In our case, it is natural to choose a Euclidean geometry
rather than geometries associated to ${\mathcal P}$  as a homogeneous
space. In particular, other choices of network constraints can lead to metric
geometry problems embedded inside Riemannian geometry problems.   For
examples, imposing sparseness on a network, or allowing for directed
edges lead to nontrivial geometries. The Euclidean average of two
sparse networks/matrices need
not be sparse, and apart from simple scalings, one expects the set
$\mathcal{L}$ of sparse matrices, properly defined, to be a discrete
subset of the manifold of positive semi-definite matrices (PSD) and
hence far from convex. Thus, it is natural to define the average of
two sparse matrices to be the sparse matrix closest to the Euclidean
average, but this may be computationally unappealing.  Moreover, the
Riemannian measure on PSD does not determine a measure on
$\mathcal{L}$, so computing Fr\'echet means becomes problematic.  Of
course, one can impose a uniform distribution on $\mathcal{L}$, but
this risks losing all geometric relations between $\mathcal{L}$ and
PSD. Hence, there are a variety of open problems to be studied examining
the implications of network structural constraints on the space
$\mathcal{L}$.  

Furthermore, since the asymptotic theory we exploit from shape
analysis relies heavily on the topological and geometrical properties of
the space within which they are brought to bear, we can expect that
different network constraints will require different levels of effort
in producing central limit theorems.  More precisely, while a general
asymptotic distribution theory for Fr\'echet means in metric spaces
has recently been derived by \citet{Bhattacharya2016}, this
theory requires that a number of conditions be satisfied, the
verification of which can be expected to become increasingly
difficult as the geometry of the space becomes complicated. Thus,
accompanying the various extensions in geometry described above are
likely to be corresponding challenges in probability theory and shape
analysis. Some progress in this direction have been spearheaded by
\citet{Bhattacharya2011} and \citet{Hotz2013}, who have considered stratified
spaces, and sticky CLTs for open books, respectively. Moreover,
similar data object analyses have been conducted using phylogenetic
trees \citep{Skwerer2013}. In object data analysis,
the approach adopted in this paper would be regarded as extrinsic, in the sense
that it embeds the manifold of interest in an ambient space. Further
research may also investigate intrinsic approaches to study the set of
graph Laplacians.

Finally, while the 1000 FCP data set is unique in its magnitude and
richness, which in turn has allowed us to pose and answer a good
number of questions relevant to neuroscience in the analyses using our
proposed testing framework, there remains much additional
empirical work to be done applying our methods, in order to more fully
establish both their capabilities and their limitations.  We would
anticipate that with the recently started BRAIN initiative, and other
endeavors like it, that within five years there will be a plethora of
databases of network-based objects in neuroscience, providing more
than ample motivation not only for the further testing of methods like
the ones we have proposed here, but also for extending
other tools from classical statistics to network data.

\section*{Acknowledgements}
The data from the 1000 Functional Connectome
Project was accessed through the 
International Neuroimaging Data-sharing Initiative (INDI),
which was designed for unrestricted data-sharing via the Neuroimaging
Informatics Tool and Resources Clearinghouse (NITRC). 
We are indebted to Sean Markan, Lizhen Lin, 
Emily Stephen and Heather Shappell for useful
suggestions and discussion. We are also very grateful for the comments
that we have received from three anonymous referees, one associate
editor, and one of the AOAS area editors.

%%%%%%%%%%%%%%%%%%%%%%%%%%%%%%%%%%%%%%%%%%%%%%%%%
%%%%%%%%%%%%%%%%%%%%%%%%%%%%%%%%%%%%%%%%%%%%%%%%%
%%%%%%%%%%%%%%%%%%%%%%%%%%%%%%%%%%%%%%%%%%%%%%%%%
\section{Supplementary Material}
We here provide detailed proofs of the main results in this paper. 

%%%%%%%%%%%%%%%%%%%%%%%%%%%%%%%%%%%%%%%%%%%%%%%%%
\noindent {\sc Proof of Theorem 1.} Let the matrix $E$ of order
$d\times d$ be partitioned in the following manner, 
$$ d-1 \ \ 1\ \ \ \ \ \ \ \ \ $$
\vskip -0.15 in 
$$E = \left(\begin{array}{cc} A &v\\ v\pri&
      x\end{array}\right)\ \ \ \begin{array}{c} d-1\\1\end{array}$$
This matrix is assumed to satisfy conditions (1), (2), and (4). 
We will call the set of such matrices $\mathcal T.$
Assume that $A$, the top left $(d-1)\times (d-1)$ block of $E$, has
nonzero determinant. We want to show that some $d(d-1)/2$-dimensional
ball around $E$  continues to lie in $\mathcal T$.
Since the rank of $E$ is $d-1$, the last column of $E$ is a linear
combination of the first $d-1$ columns.  Since the columns of $E$ add
to zero and $E$ is symmetric, the rows of $E$ add to zero.  For
$v\pri = (v_1,\ldots, v_{d-1}),$ we must have 
$$v_i = -\sum_{j=1}^{d-1} A_{ij}, \qquad\te{and}\qquad x =
-\sum_{j=1}^{d-1} v_j.$$ Thus, $v$ and $x$ are determined by the
entries of $A$.  

The matrix, $A$, is symmetric. Thus, it lies in the subspace $S$ of
$\R^{(d-1)^2}$  of dimension $d(d-1)/2$ consisting of symmetric matrices.
${\rm Det}(A)\neq 0$, so for some matrix $A_{\ep}$ in some small neighborhood $U$ of
$A$ in $S$, ${\rm det}(A_{\ep}) \neq 0.$  Each choice of $A_{\ep}$ determines
a corresponding $v$ and $x$.  Conversely, each $E_{\ep}\in \mathcal
T$ sufficiently close to $E$ in the $\R^{d^2}$ norm has $\det(A_{\ep}) \neq
0$ and $A_{\ep}-A = X$ is symmetric, so $A_{\ep}$ and hence $E_{\ep}$ is determined by $X$.
Thus, a neighborhood of $E$ in $\mathcal T$ is bijective to $U$. It is
easy to check that this bijection is a diffeomorphism.

If some other $(d-1)\times (d-1)$ block $B$ of $E$ 
has nonzero determinant, we note that the top $(d-1)\times (d-1)$ block 
$A$ of the matrix determines the entire matrix as above. Any small
symmetric perturbation $A_{\ep}$ of $A$ (with the necessary
perturbations of the last row and column to preserve (4)) still satisfies 
$\det(B_{\ep}) \neq 0$. Conversely, any  $B\in \mathcal T$
sufficiently close to $E$ so that $\det(B_{\ep}) \neq 0$ determines a
symmetric perturbation of $A$ as above.
Hence, we again obtain a neighborhood of $E$ in $\mathcal T$ parametrized 
by a neighborhood $U$ of $A$ in $S$. This shows that  $\mathcal T$ is
a submanifold of $\R^{d^2}$ of dimension $d(d-1)/2.$  
The set of matrices satisfying (5) alone is an open convex cone in $\R^{d^2}$. 
When we intersect the submanifold $\mathcal T$ with this cone, we get
an open submanifold $\mathcal T'$ of $\mathcal T$.  Thus
$\mathcal T'$, the set of matrices with (1), (2), (4), (5),  is also a
submanifold of $\R^{d^2}$ of dimension $d(d-1)/2.$  

The space $\mathcal T'$ has several connected components.  A matrix $E_0$ with 
$k$ positive eigenvalues and a matrix $E_1$ with $k'\neq k$ positive
eigenvalues lie in different components, as a path in $\mathcal T'$
from $E_0$ to $E_1$ would contain a matrix with a zero eigenspace of
multiplicity at least two.  Conversely, if $k=k'$, then $E_0$ and
$E_1$ are in the same component of $\mathcal T'$.  For 
the line segment $E_t = (1- t)E_0+ tE_1$ stays in
$\mathcal T'$, for every $t\in [0,1]$. Since the
components are open, the component of $\mathcal T'$ satisfying $k =
d-1$ is again a submanifold of dimension $d(d-1)/2.$  But this
component has condition (3), and so is precisely $\cL_d.$  This proves
that $\cL_d$ is a manifold of dimension $d(d-1)/2$. 

For the convexity statement, conditions (2) -- (5) are convex conditions; e.g. for (3), if
$A$ and $B$ are positive semidefinite, then
$$\langle (tA + (1-t)B)v,v\rangle = t\langle Av,v\rangle +
(1-t)\langle Bv,v\rangle \geq 0$$
for $t\in [0,1]$ and $v\neq 0$.  Clearly, (1) -- (5)
together is a convex condition.  For if $A$ and $B$ satisfy (1) --
(5), then $A$ and $B$ come from weighted connected graphs, as does
$tA+ (1-t)B.$  Since a graph is connected iff the rank of the
corresponding Laplacian matrix has rank $d-1$, the rank of 
$tA+ (1-t)B$  is $d-1$ for $t\in [0,1].$
 Thus $\cL_d$ is a convex submanifold of $\R^{d^2}.$  
 
To show that $\cL_d$ lies in an affine subset, fix $E\in
 \cL_d.$ For $k = d(d-1)/2$, take $k$ distinct points $s_i$ in 
 $\cL_d$, none of them equal to $E$, such that the convex hull of
 these points contains $E$.  (For example, two of the points can be
 close to  $E\pm S$ for a small symmetric matrix $S$.) For generic
 choices, the $k$ points plus $E$ determine an (affine) $k$-plane $P$,
 and the convex  hull of these points lies in both $P$ and $\mathcal
 S$.  Since $P$ and $\cL_d$ have the same dimension, the open
 convex hull is exactly a neighborhood of $E$ in $\cL_d$.  
 
 We now show that the plane $P$ is independent of the choice of $E.$
 Since $\cL_d$ is convex, it is connected.  Take $F\in \mathcal
 S$, let $\ell$ be the Euclidean line segment from $E$ to $F$,  and
 set $E_t = (1-t)E + tF\in \cL_d.$  Arguing as above, we find a
 plane $P_t$ containing a neighborhood $V_t$ of $E_t$ in $\cL_d.$
 By compactness, there exist 
 $0 = t_0,\ldots, t_n = 1$ with $\cup_{i=0}^n V_{t_i} \supset \ell.$
 If $P = P_0 \neq P_{t_1}$,  then some line segment from one of the $s_i$'s
 determining $P_0$ to one of the $s_j$'s determining $P_{t_1}$ does not
 lie in $\cL_d$, a contradiction. Thus $P = P_{t_1}$, and by induction,
 $P= P_1.$ Since $F$ is arbitrary in $\cL_d$, it follows that
 $\cL_d$ %and hence $ \cL_d'$ 
 lies in $P$.  \hfill $\Box$

\bigskip

\noindent {\sc Proof of Corollary 1.} 
In the notation of the proof of Theorem 1, assume that $E$ has
conditions (1), (2), (4), (5${}'$).  Then $A$ is symmetric and has
$a_{ij}\leq 0.$  Thus $A$ is in bijection with the closed ``quadrant''
$\{(x^1,\ldots, x^{d(d-1)/2}): x^i\leq 0\}$, which is the basic
example of a manifold with corners.  If the rank $d-1$ submatrix $B$
of $E$ is not in the top left corner, a relabeling of coordinates
moves $B$ to the top left corner.  Since the relabeling takes the
closed quadrant to a closed quadrant, a neighborhood of $B$ has the
structure of a manifold with corners.  It is trivial to check that
transition maps from chart to chart are smooth.  If we impose (3),
then as in the previous proof we pick out one connected component of this
manifold with corners, and each component is a manifold with corners.
The statements on convexity and affine subspaces follow immediately from Theorem
1, since $\cL'_d$ is a dense subset of $\cL_d$. \hfill $\Box$
%and that $a_{ij} \leq 0$.  The set of all such $A$ 
% is the closure of the positive ``quadrant" in $\R^{n(n-1)/2}$ and so
% is a manifold with corners.

\bigskip

\noindent {\sc Proof of Theorem 2.} Assume the $\ell\times \ell$ block
with nonzero determinant occurs in the top left corner;
the other cases are handled as in the proof of Theorem 1.  Thus let
%\vfill\eject
$$ \ell \ \ \ \ \ \ \ d-\ell\ \ \ \ \ \ \ \ \ $$
\vskip -0.15 in 
$$E = \left(\begin{array}{c|ccc} A &v_1&\ldots&v_{d-\ell}\\ \hline
v_1\pri&&&\\
\vdots& b_1&\ldots& b_{d-\ell}\\
v_{d-\ell}\pri&&&\end{array}\right)\ \ \ \begin{array}{ccccc}\ell\\
\\d-\ell\\ \\  \end{array}$$

\noindent have conditions (1${}_\ell$), (2), (4). 
 Here, $v_i$ is an $\ell\times 1$ column vector, and $ b_i$ is
 a $(d-\ell)\times 1$ column vector.  The dimension of the set of
 $\ell\times \ell$ symmetric matrices $A$ with nonzero determinant is 
 $\ell(\ell+1)/2.$  Since the last $d-\ell$ columns must be linear
 combinations of the first $\ell$ columns, we have
 $$v_i = \sum_{j=1}^\ell v_{ij} a_j,\qquad  i \in \{1,\ldots, d-\ell\};$$
 where $ a_j$ is the j${}^{\rm th}$ column of $A$.  The $v_{ij}$'s
 are arbitrary for $i = 1, \ldots, d-\ell -1,$ 
but (4) implies that the $v_{d-\ell, j}$'s are determined by the
previous $v_{ij}$'s. Therefore, we get another  
$(d-\ell -1)\ell$ degrees of freedom (i.e. dimensions), so the
dimension of the space of matrices with 
(1${}_\ell$), (2), (4) is $\ell(\ell+1)/2 + (d-\ell-1)\ell =
d\ell - \ell(\ell+1)/2$.  The argument for adding in
conditions (3) and (5) goes as before. \hfill $\Box$

%%%%%%%%%%%%%%%%%%%%%%%%%%%%%%%%%%%%%%%%%%%%%%%%%%
\noindent {\sc Proof of Theorem 3.} 
The Laplacian CLT considered in this paper is a specialization of a
general result due to \citet{Bhattacharya2016}, which considers a
metric space $(\cX,\rho)$ equipped with a probability measure $Q$.
In addition to the conditions stated in the main body of the paper,
two further regularity assumptions must be made on the first and
second derivatives of the function $\rho^{2}(\phi^{-1}(u),x)$. These
conditions are described below as (A5) and (A6). 

\citet{Bhattacharya2016} have shown that Euclidean coordinates of a
Fr\'echet mean defined on a metric space converges to a normal
distribution, under the following assumptions:
(A1) the Fr\'echet mean $\mu$, as described in equation
(1) is unique; (A2) $\mu\in
A\subseteq\cX$, where $A$ is $Q$-measurable, and $\wh{\mu}_{n}\in A$,
almost surely; (A3) there exists a
homeomorphism $\phi:A\to U$, for some $s\geq1$, where $ U$ is an
open subset of $\R^{s}$; (A4) for every $u\in U$, the map, $u
\mapsto h(u;x) := \rho^{2}(\phi^{-1}(u),x)$, is twice differentiable
on $ U$, for every $x\in\cX$ outside a $Q$-null set; (A5) for every
pair $1\leq k,l\leq s$, with $u\in U\subseteq\R^{s}$ and $x\in\cX$, letting
\begin{equation}\notag
       D_{k}h(u;x) := \frac{\partial}{\partial u_{k}} h(u;x),
       \qquad\te{and}\qquad
       D_{k,l}h(u;x) := \frac{\partial^{2}}{\partial u_{k}\partial u_{l}} h(u;x),
\end{equation}
we require that $\E\lt[|D_{k}h(u;x)|^{2}\rt] <\infty$, and
$\E\lt[|D_{k,l}h(u;x)|\rt] <\infty$; moreover, (A6) defining
$f_{k,l}(\ep,x):=\sup\{|D_{k,l}h(u;x) - D_{k,l}h(\phi(\mu);x)|:
|u-\phi(\mu)|<\ep\}$, we also require modulus continuity, such that 
$\E[|f_{k,l}(\ep;Y)|] \to 0$, as $\ep\to 0$, for every $1\leq k,l \leq s$;
and finally, (A7) the matrix, $B:=
\{\E[D_{k,l}h(\phi(\mu);Y)]\}_{k,l=1,\ldots,s}$, should be
non-singular. Under these conditions, it is then true that the
following convergence in distribution holds,
\begin{equation}\notag
      n^{1/2}\lt(\phi(\wh{\mu}_{n}) - \phi(\mu)\rt) \longrightarrow
      N(0,B^{-1}VB^{-T}),
\end{equation}
where $V:=\cov[D\,h(\phi(\mu);Y)]$ is assumed to be non-singular. 

In our setting, we have drawn an iid sample of combinatorial Laplacians
from an unknown generating distribution, such that we have $Y_{i}
\sim F(\La,\Sig)$, for every $i=1,\ldots,n$, 
where $\La$ and $\Sig$ are the mean Laplacian and the covariance matrix of the
upper triangle of $Y$, with respect to some unknown distribution,
$F$. Observe that the space of interest is here $\cL\pri_{d}$,
equipped with the Frobenius distance, as stated in Corollary
1, thereby forming the metric space,
$(\cL\pri_{d},\|\cdot\|_{F})$. We will see that conditions (A1) -- (A4) as
well as (A7) are necessarily satisfied in our context. Moreover, we
will assume that conditions (A5) and (A6) also hold. 

Condition (A1) is readily satisfied, since we have demonstrated that
the space of interest, $\cL\pri_{d}$, is a convex subspace of $\R^{d^{2}}$; and
moreover the arithmetic mean is a convex function on that space by
Corollary 1. Thus, the sample Fr\'echet mean, 
$\wh{L}_{n}$, is unique, for every $n\in\N$. Secondly, we have assumed
that the underlying measure gives a non-zero positive probability to a
subset $U\in\R^{d^{2}}$, which contains $\La$. Therefore, condition (A2) is
satisfied, in the sense, that there exists a subset
$A\subseteq\M_{d,d}(\R^{+})$, such that $A$ is
$\p$-measurable. In addition, since the strong law of large numbers
holds for the Fr\'echet mean \citep[see][]{Ziezold1977}, 
we also know that $\wh{L}_{n}\to\La$, almost surely;
and therefore, $\p[\wh{L}_{n}\in A] \to 1$, as $n\to\infty$, as
required by condition (A2). 

For condition (A3), observe that, in our context, the homeomorphism of interest,
$\phi:A\mapsto U$, is the \textit{half-vectorization}
function. This takes a matrix in $\cL\pri_{d}$, and returns a vector
in $\R^{\binom{d}{2}}$, such that for every $Y\in\cL\pri_{d}$, $\phi(Y) :=
\vech(Y)$. Specifically, this vectorization is defined by a change of
indices, such that for every $i\leq j$, with $1\leq i,j\leq d$, we
have $[\phi(Y)]_{k(i,j)} := y_{ij}$, with $k(i,j):=(i-1)d + j$.
The inverse function, $\phi^{-1}$, is then readily obtained for every
$u\in U\subseteq\R^{\binom{d}{2}}$, satisfying $\phi^{-1}(u)=Y$, as
$[\phi^{-1}(u)]_{ij} = y_{ij}$. The bicontinuity of $\phi$ is hence
trivially verified and this map is therefore a homeomorphism.

For condition (A4), the function $h(u;Y):=\rho^{2}(\phi^{-1}(u),Y)$,
for every $u\in U\subseteq\R^{\binom{d}{2}}$ and every $Y\in\cL\pri_{d}$,
outside of a $Q$-null set, is here defined as 
\begin{equation}\notag
    h(u;Y) := ||\phi^{-1}(u)- Y||_{F}^{2} = \sum_{i\leq j}^{d}\,
              \Big([\phi^{-1}(u)]_{ij} - y_{ij}\Big)^{2},
\end{equation}
where the sum is taken over all the pairs of indices $1\leq i,j\leq
d$, satisfying $i\leq j$. The first derivative of this map
with respect to the coordinates of the elements of $\cL\pri_{d}$ in
$\R^{\binom{d}{2}}$, is straightforwardly obtained. Setting
$X:=\phi^{-1}(u)$, we have
\begin{equation}\notag
      D_{k(i,j)}h(u;Y) := 
       \frac{\partial}{\partial u_{k(i,j)}} ||\phi^{-1}(u)- Y||_{F}^{2} 
       = 2(x_{ij} - y_{ij}). 
\end{equation}
The second derivative of $h(u;Y)$ can be similarly derived for every
quadruple, $1\leq i,j,i\pri,j\pri\leq d$, satisfying $k(i,j)\neq k(i\pri,j\pri)$.
When expressed with respect to $\La\in U$, this gives 
\begin{equation}\notag
    D_{k(i,j),k(i\pri,j\pri)}h(\phi(\La);Y) = 
    \begin{cases}
       2, & \te{if } k(i,j) = k(i\pri,j\pri), \\
       0, & \te{otherwise}.
    \end{cases}
\end{equation}
It immediately follows that the matrix of second derivatives
is $B=2I$, and hence condition (A4) is verified. In addition, we have
assumed that conditions (A5) and (A6) hold in our context.
Finally, we have seen that the matrix $B$ is diagonal and hence
non-singular, as required by condition (A7).

We can also compute the covariance matrix of the
resulting multivariate normal distribution. For this, we require the 
matrix $V:=\cov[D\,h(\phi(\La);Y)]$. Given our choice of $\phi$, we
need to consider the mean vector of $D\,h(\phi(\La);Y)$, which is given
for every $1\leq i,j\leq n$ by $\E[D_{k(i,j)}\,h(\phi(\La);Y)] =
2(\La_{ij} - \E[Y]_{ij}) = 0$. We can then compute the elements of
$V$. For every quadruple $1\leq i,j,i\pri,j\pri\leq n$, this gives
\begin{equation}\notag
  \begin{aligned}
   V_{k(i,j),k(i\pri,j\pri)} 
    &= \E[D_{k(i,j)}\,h(\phi(\La);Y)\cdot D_{k(i\pri,j\pri)}\,h(\phi(\La);Y)] \\
    &= 4\E[(\La_{ij} - Y_{ij})(\La_{i\pri,j\pri} - Y_{i\pri j\pri})] \\
    &= 4\big(\E[Y_{ij}Y_{i\pri j\pri}]-\La_{ij}\La_{i\pri,j\pri}\big),
  \end{aligned}
\end{equation}
since the cross-term vanishes, after taking the
expectation. Therefore, the asymptotic covariance matrix in Theorem
3 is indeed equal to the covariance matrix of the
distribution, from which the $Y_{i}$'s have been sampled. That
is, this covariance matrix is given by
$B^{-1}VB^{-T} = (2I)^{-1} V (2I)^{-1} = \var[\phi(Y)] = \Sig$.
Therefore, all the conditions of Theorem 2.1 of 
\citet{Bhattacharya2016} have been satisfied, and hence
$n^{1/2}(\phi(\wh{L}_{n}) - \phi(\La)) \to N(0,\Sig)$, as stated in
Theorem 3.

% references --------------------------------------------------
\bibliographystyle{imsart-nameyear}
\bibliography{/home/cgineste/ref/bibtex/Statistics,%
              /home/cgineste/ref/bibtex/Neuroscience,%
              refs-SR}
\end{document}